\documentclass[12pt,a4paper,english]{article}
\usepackage[a4paper, margin=2.5cm]{geometry}

\usepackage[utf8]{inputenc}
\usepackage[T1]{fontenc}
\usepackage[english]{babel}

\usepackage{authblk,amsmath,amsfonts,amssymb,amsthm,amsbsy,bm,calrsfs,graphicx,float}
\usepackage{dcolumn}
\usepackage{color,xcolor}
\usepackage{soul}
\usepackage{nicefrac}
\usepackage[colorlinks=true,citecolor=blue]{hyperref}
\usepackage[font=small,labelfont=bf]{caption}
\usepackage{tikz}
\usepackage{stackengine}
\usepackage{subfig}
\usepackage{hyperref}
\usepackage{cleveref}
\usepackage{comment}
\usepackage{gensymb}
\usepackage{siunitx} 
\usetikzlibrary{decorations.pathmorphing}

\usepackage[shortlabels]{enumitem}

\newcommand{\gammaas}{\gamma_{\rm as}}
\newcommand{\gammaal}{\gamma_{\rm a\ell}}
\newcommand{\gammals}{\gamma_{\rm \ell s}}
\newcommand{\lambdacap}{\lambda_\mathrm{c}}

\newcommand{\bs}[1]{\boldsymbol{#1}}
\newcommand{\bF}{\boldsymbol{F}}
\newcommand{\Gstiff}{G_{\rm SG184}}
\newcommand{\Gsoft}{G_{\rm SG186}}

\usepackage{academicons}
\definecolor{orcidlogocol}{HTML}{A6CE39}

\makeatletter
\def\@fnsymbol#1{\ensuremath{\ifcase#1\or *\or \ddagger\or
\mathsection\or \mathparagraph\or \|\or **\or \dagger\dagger
\or \ddagger\ddagger \else\@ctrerr\fi}}
\makeatother

\graphicspath{{./pics}}

\title{Shape of Polystyrene Droplets on Soft PDMS: Exploring the Gap Between Theory and Experiment at the Three-Phase Contact Line}

\author[1]{Khalil Remini\thanks{k.remini@physik.uni-saarland.de}}
\author[2]{Leonie Schmeller\thanks{leonie.schmeller@wias-berlin.de}}
\author[2]{Dirk Peschka\thanks{dirk.peschka@wias-berlin.de}}
\author[2]{Barbara Wagner\thanks{barbara.wagner@wias-berlin.de}}
\author[1]{Ralf Seemann\thanks{r.seemann@physik.uni-saarland.de}}

\affil[1]{\small Department of Experimental Physics, Saarland University, 66123 Saarbrücken, Germany}
\affil[2]{\small Weierstrass Institute for Applied Analysis and Stochastics, 10117 Berlin, Germany}
\date{\today}

\begin{document}
\maketitle

\begin{abstract}
The shapes of liquid polystyrene (PS) droplets on  viscoelastic polydimethylsiloxane (PDMS) substrates are investigated experimentally using atomic force microscopy for a range of droplet sizes and substrate elasticities. These shapes, which comprise the PS-air, PS-PDMS, and PDMS-air interfaces as well as the three-phase contact line, are compared to theoretical predictions using axisymmetric sharp-interface models derived through energy minimization. We find that the polystyrene droplets are cloaked by a thin layer of uncrosslinked molecules migrating from the PDMS substrate. By incorporating the effects of cloaking into the surface energies in our theoretical model, we show that the global features of the experimental droplet shapes are in excellent quantitative agreement for all droplet sizes and substrate elasticities.
However, our comparisons also reveal systematic discrepancies between the experimental results and the theoretical predictions in the vicinity of the three-phase contact line. Moreover, the relative importance of these discrepancies systematically increases for softer substrates and smaller droplets. We demonstrate that global variations in system parameters, such as surface tension and elastic shear moduli, cannot explain these differences but instead point to a locally larger elastocapillary length, whose possible origin is discussed thoroughly.
\end{abstract}

\section{Introduction}
%
%
Compared to the rather mature understanding of wetting and dewetting phenomena on rigid solid substrates, the physics of wetting on soft adaptive substrates lags behind due to a number of coupled physical phenomena that are involved. 
This includes the conceptual difference between surface tension and the free surface energy for solid surfaces that was pointed out in the pioneering work of Shuttleworth \cite{Shuttleworth-1st}, the dissipation in a microscopic \emph{wetting ridge} that can influence the macroscopic dynamics by \emph{viscoelastic braking}, which was introduced by Carr{\'e} et al. \cite{Shanahan-1st}, and the time-dependent poroelastic relaxation of the soft solids \cite{Xu2020}.
It is only in recent years that an increasing number of theoretical and experimental studies have addressed the complex nature of these adaptive processes, in particular in the vicinity of a moving three-phase contact line (TPCL), e.g. \cite{style2017elastocapillarity,bico2018elastocapillarity,chen2018static,andreotti2020statics}.

The most important quantity that governs classical elastic wetting near a TPCL is the elastocapillary length
\begin{align}
\label{eqn:elastocap}
\lambdacap = \frac{\gamma}{G},
\end{align}
measuring the length scale, below which usually capillarity effects dominate over elastic effects, where $\gamma$ is the surface tension {of the wetting fluid} and $G$ the elastic shear modulus {of the soft solid}. 
For soft wetting on viscoelastic substrates, different regimes exist depending on the relative magnitude of the elastocapillary length compared with molecular scales $a \sim 10^{-9}\,\si{\meter}$, droplet size $R$, and substrate thickness $H$. 

\begin{figure}[b!]
\centering

{\includegraphics[width=0.31\textwidth,trim={0cm 30cm 0cm 20cm},clip]{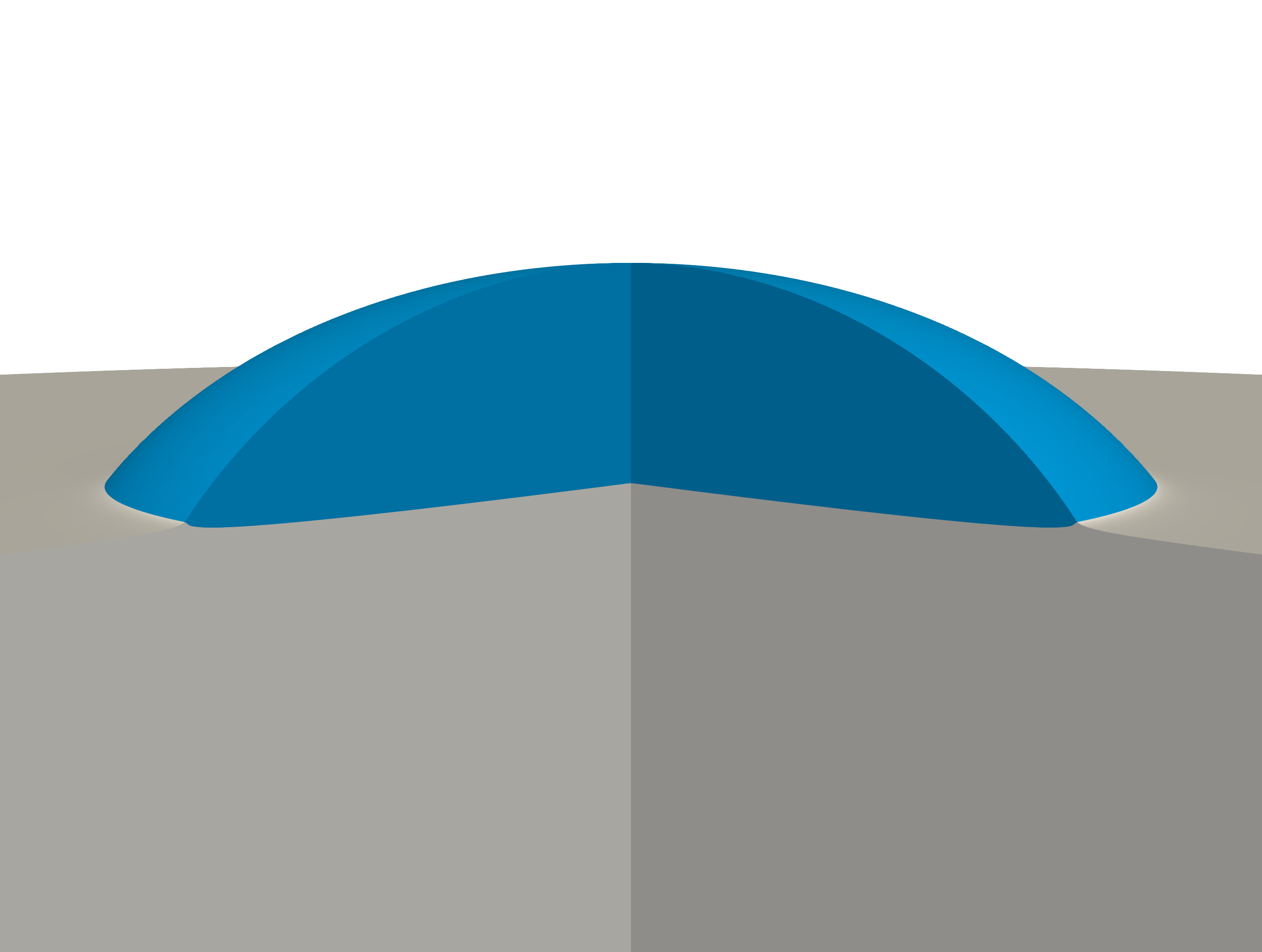}}\hfill
{\includegraphics[width=0.31\textwidth,trim={0cm 30cm 0cm 20cm},clip]{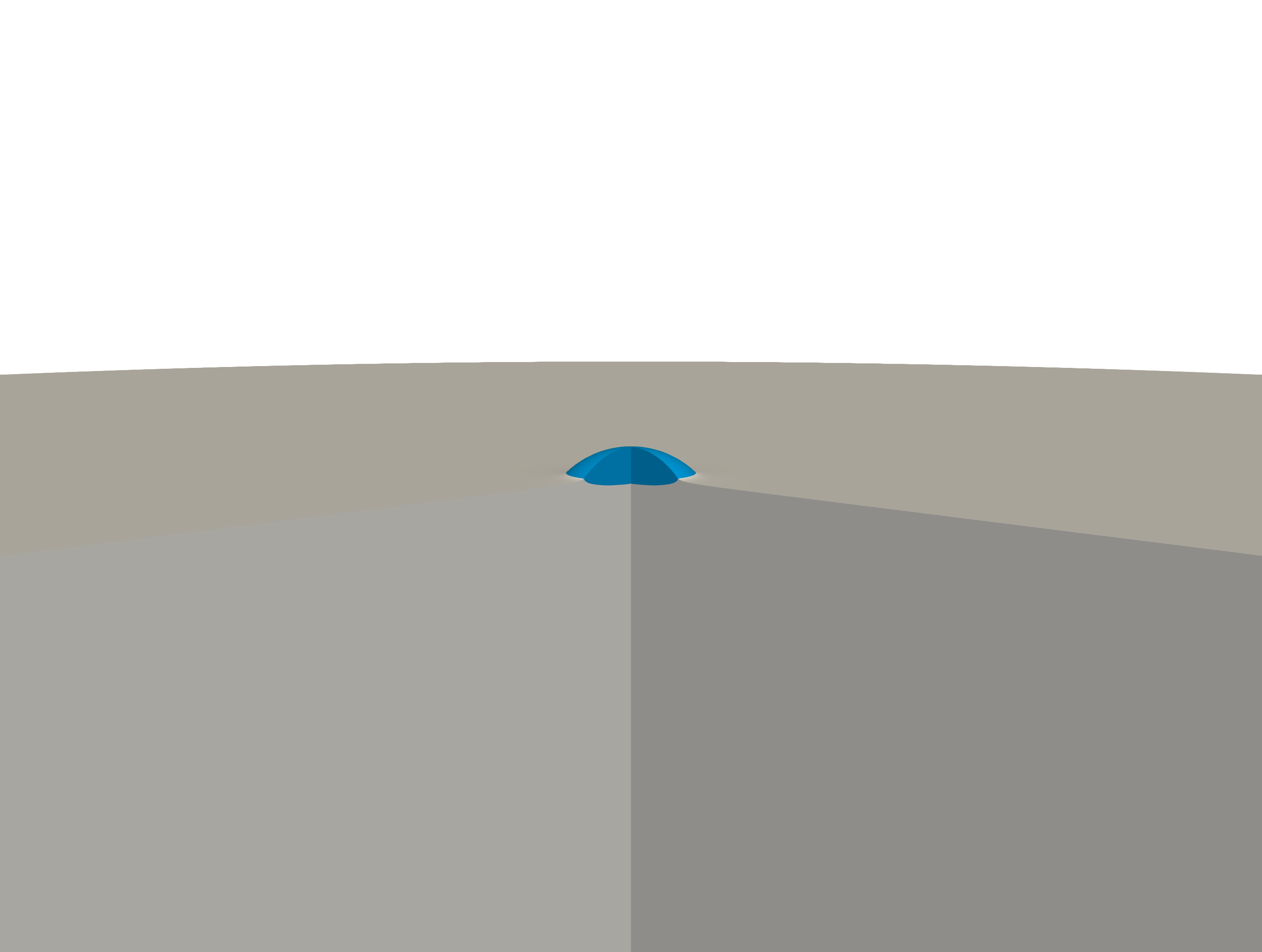}}\hfill
{\includegraphics[width=0.31\textwidth,trim={0cm 30cm 0cm 20cm},clip]{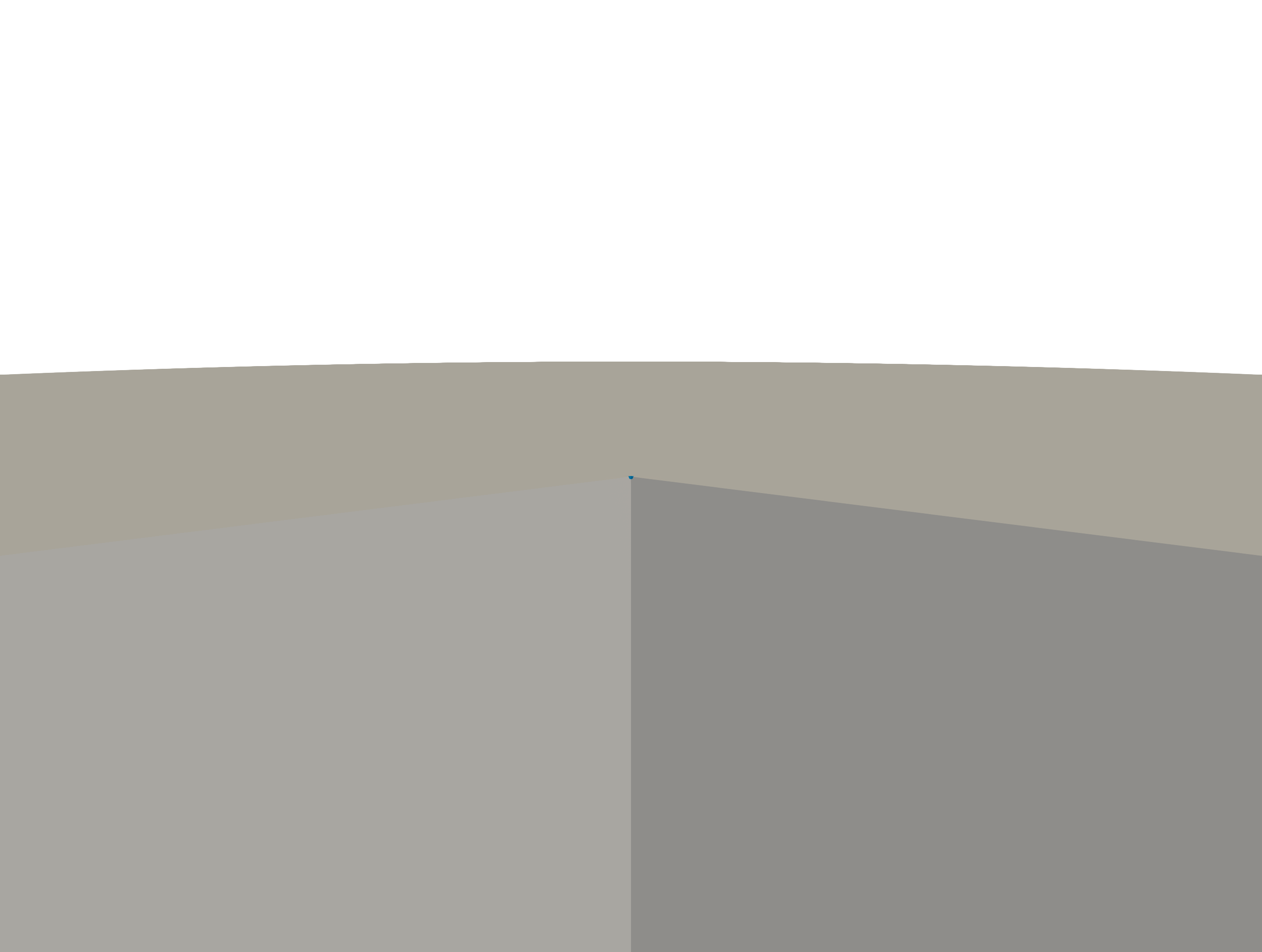}}

\vspace*{0.3cm}

{\includegraphics[width=0.31\textwidth,trim={0cm 0cm 0cm 20cm},clip]{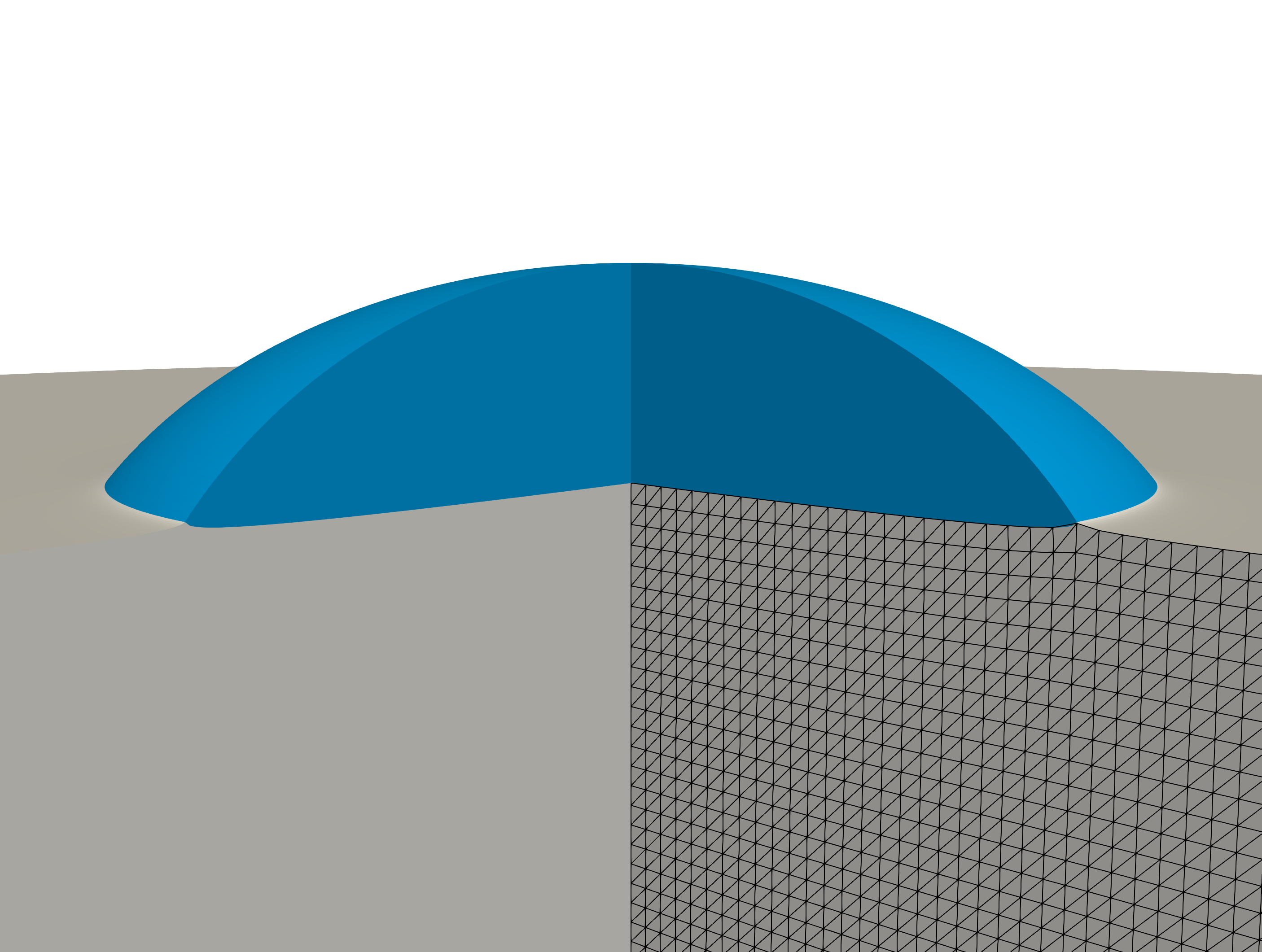}}\hfill
{\includegraphics[width=0.31\textwidth,trim={0cm 0cm 0cm 20cm},clip]{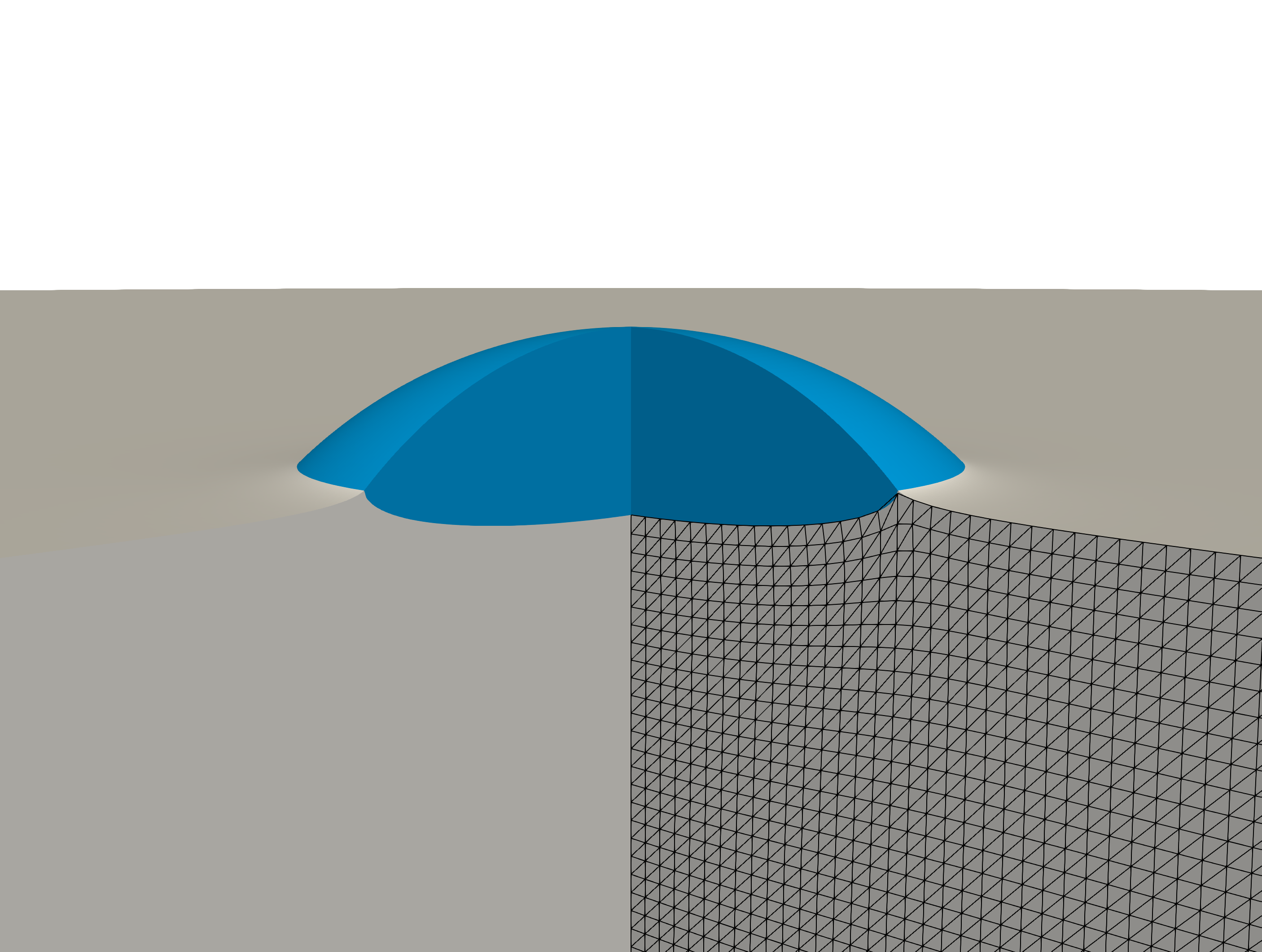}}\hfill
{\includegraphics[width=0.31\textwidth,trim={0cm 0cm 0cm 20cm},clip]{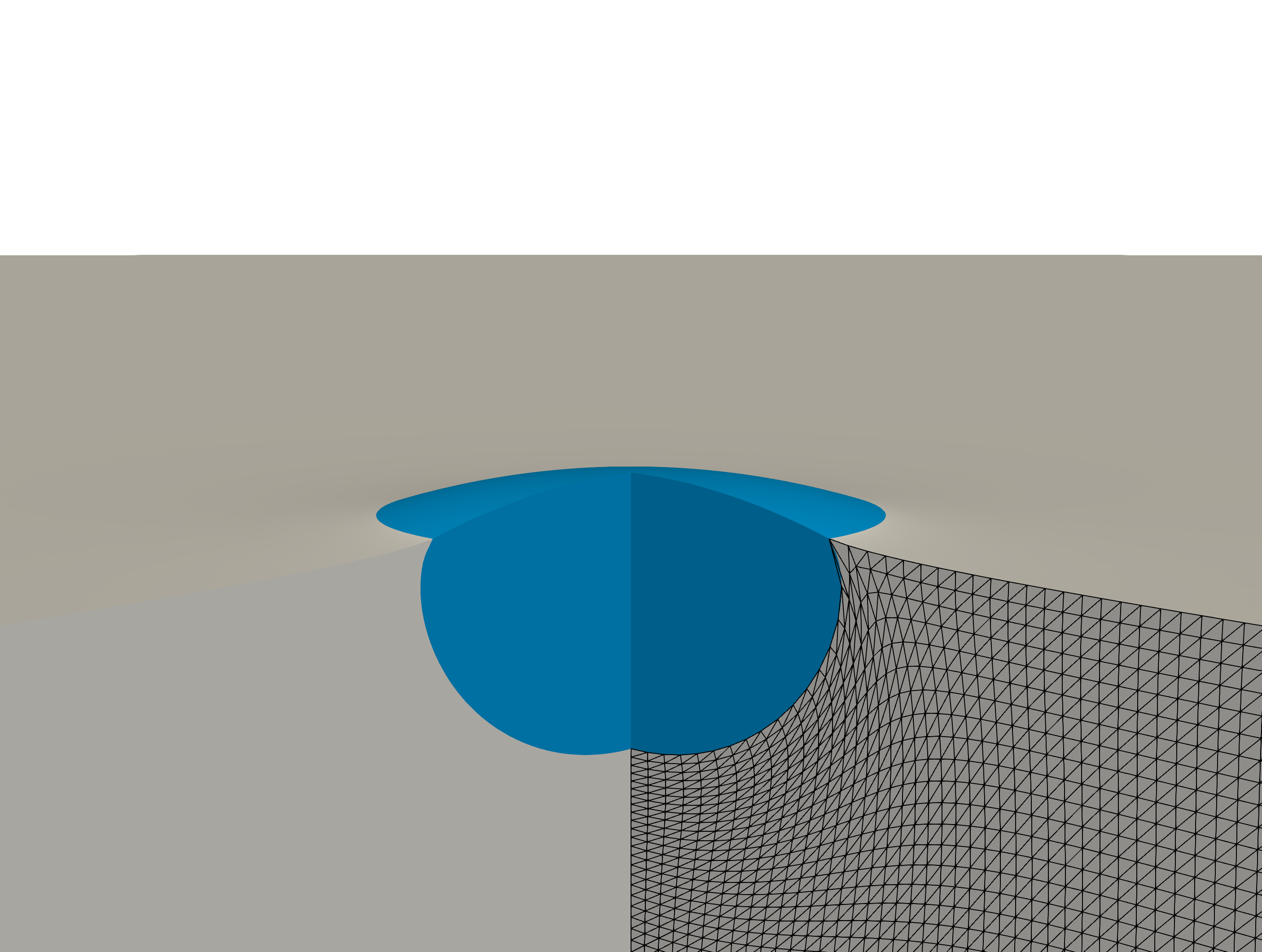}}

\caption{Axisymmetric stationary liquid droplets (blue) on an elastic substrate (gray) with height $H = 7~\si{\micro\meter}$, computed using finite element simulations. The droplet radii are (left) $R \approx 2.5~\si{\micro\meter}$, (middle) $R \approx 300~\si{\nano\meter}$, and (right) $R \approx 13~\si{\nano\meter}$, based on SG186 material parameters with a capillary length $\lambdacap = 80~\si{\nano\meter}$. The upper panel shows all droplets at a fixed scale of $3~\si{\micro\meter}$ to highlight their relative sizes. The black mesh illustrates elastic deformations by depicting the deformation of the substrate material reference frame and highlighting the singularity at the TPCL. This mesh differs from the finer computational mesh used in simulations.}
\label{fig:theoretical_droplets_intro}
\end{figure}

For length scales below $a$, the interfaces can be understood as diffuse and interface forces are smoothed out over corresponding distances. 
On length scales above $a$, we distinguish the following regimes: 
(i) the rigid substrate $\lambdacap\ll a \ll R$, (ii) the moderately soft substrate $a\ll\lambdacap\ll R$  and (iii) the soft limit $a\ll R\ll \lambdacap$, e.g.~\cite{andreotti2020statics}. The rigid limit (i) is solely governed by Young's law \cite{young1805} and substrate deformations are practically invisible. The case (ii) still allows for Young's law on the macroscopic scale, but satisfies Neumann's law  near the TPCL \cite{neumann1894vorlesungen}, where an elastic ridge becomes detectable. In the soft limit (iii) elastic properties can be mainly neglected and stationary droplets are liquid lenses determined by the Neumann triangle. For a more detailed discussion, including the regimes of thin, thick, and semi-infinite substrates, we refer to \cite{style2012static}.
Typical droplets for various radii on moderately thick substrates $H/R>1$ obtained by minimization of elastic  and surface energy are shown in \Cref{fig:theoretical_droplets_intro}.
Experimental measurements of the elastocapillary ridge for moderately soft but also thinner substrates $H=50\,\si{\micro\meter}$ can be found in \cite{style2013universal} for $\lambdacap\approx 10\,\si{\micro\meter}$ and droplet radii in the range $27\,\si{\micro\meter}<R<226\,\si{\micro\meter}$. 

%
The Shuttleworth effect relates the surface tension to the surface energy  depending on a (specific) surface area, where for one-component liquids they are assumed to be equal \cite{Shuttleworth-1st}. This relation implies that for solid surfaces, surface tension and surface energy are different concepts. While experimental evidence for a positive Shuttleworth effect for PDMS samples was presented in \cite{style2013universal,Xu2017,Park2014,snoeijer2018paradox,Bain2021,Heyden2021,Zhao2022} it was not observed in \cite{Schulman2018} and argued that no Shuttleworth effect is expected in case of a polymer with reduced cross linking density at the interface. So the resulting role and interpretation of the Shuttleworth effect has been slightly controversial. 
Despite this controversy, it is generally accepted that 
force balances at a TPCL are derived using variational energetic arguments as in \cite{henkel2021gradient} or \cite{snoeijer2018paradox,pandey2020singular}, where the former employs dimension-reduced models with disjoining pressures and the latter two show that in addition to Neumann's law another condition is derived and ensures a \emph{no-pinning} condition. 

%
Beside the balance of elastic and surface energies, there are a couple of irreversible phenomena that emerge from the viscoelastic or poroelastic nature of soft substrates. It is well known that stationary droplets can generate substantial stresses that lead to crack formation (fracture) in underlying soft substrates \cite{PhysRevE.88.042410}. In dynamic wetting of soft substrates at sufficiently high velocity, contact lines do not move continuously but in a series of stick-slip events, e.g. \cite{pu2008characterization,Cy52-276_G_Karpitschka}, that are connected to the viscoelastic nature of the substrate encoded in the ratio of loss and storage modulus. Similar to Navier-slip, this effect is based on a characterization of an effective interfacial dissipation similar to Navier-slip observed in viscoelastic adhesion, where such a stick-slip process induces the rupture of adhesive bonds \cite{zhang1997effect,wu2007stick} in the vicinity of a TPCL. The distinction between viscoelastic and poroelastic effects is slightly ambiguous, but \cite{Xu2020} argue that both are essential to fully describe the relaxation dynamics of soft polymeric gels.

Upon realizing that soft substrates can be inhomogeneous materials, such as swollen aqueous hydrogels or polymer melt networks with uncrosslinked chains that can be transported within the substrate by elasto-chemo-capillary forces \cite{Style2018PRX}, it becomes evident that these transport processes can significantly alter dynamic properties. Examples include the temporal formation of a wetting ridge \cite{Zhao2018} or the slow relaxation of a wetting ridge formed at the TPCL \cite{Xu2020}, i.e., towards or away from regions of highest strain. These transport phenomena are particularly pronounced in strongly swollen networks, where it has been shown that uncrosslinked polymer phase-separates from the network near the TPCL, influencing the adhesive properties of the gel \cite{PhaseSeparationKJensen}. The same effect has been observed in static and dynamic wetting ridges of water droplets on swollen PDMS substrates \cite{PhaseSeparationJPham, hauer2023phase}.
To the best of our knowledge, a fully coupled model that integrates viscoelastic and poroelastic effects for wetting problems has yet to be developed.

In systems with multiple surface tensions $\gamma_{ij}$, stability conditions govern the existence and stability of stationary droplet configurations and of contact angle force balances. 
In particular for slippery surfaces, \emph{cloaked droplets} that are completely wetted by a lubricant exist \cite{smith2013droplet}.
A corresponding dynamical transition to cloaked droplets has been shown theoretically using  molecular dynamic simulations in \cite{CloakingHauerVollmer2022,zheng2023theoretical}. Depending on droplets being completely or partially cloaked, this effect can modify the effective surface tensions and thereby change the observed wetting dynamics \cite{style2013universal,Zhao2022,PhaseSeparationJPham}; it can also lubricate the droplets \cite{Hourlier-Fargette2017,Wong2020} increasing the velocity of sliding drops or reducing wetting hysteresis. 

In this study, we investigate stationary polystyrene (PS) droplets on two types of polydimethylsiloxane (PDMS) substrates (Sylgard 184 (SG184) and Sylgard 186 (SG186)) in the moderately soft regime, where the elastocapillary lengths are $\lambdacap \sim 30\ \si{\nano\meter}$ and $\lambdacap \sim 80\ \si{\nano\meter}$, respectively. The radii of the considered droplets range between $300 \ \si{\nano\meter}$ and $3 \ \si{\micro\meter}$. Using Atomic Force Microscopy (AFM), we directly obtain the topographies of the PS-air and PDMS-air interfaces with superior resolution compared to typical optical or X-ray methods. To expose the internal PS-PDMS interface, we employ a novel lift-off technique combined with AFM.
We rigorously compare these experimental measurements with theoretical predictions of axisymmetric droplet shapes, based on nonlinear neo-Hookean elasticity where capillary surfaces are modeled with constant surface tension. These comparisons emphasize droplet sizes and features near the three-phase contact line (TPCL), which are challenging to resolve with conventional optical or X-ray techniques. To address observed discrepancies near the TPCL, we consider the relevance of nonlinear effects, including viscoelasticity, poroelasticity, the Shuttleworth effect, cloaking, and potential phase separation—an effect not previously identified for un-swollen PDMS substrates. While these nonlinear effects are often treated separately in the literature, they may act together to influence the observed phenomena.


\section{Experimental Methods and Theoretical Models}
\label{sec:methods}
\subsection{Experimental Methods}

As support for the PDMS samples, silicon wafer cuts ($\langle 100 \rangle$, $375~\si{\micro\meter}$, Si-mat) about $(1 \times 1)~\si{\centi\meter\squared}$ were cleaned by sonication in ethanol, isopropanol, and toluene, followed by a $30~\si{\second}$ treatment in a plasma cleaner (Diener electronic Femto). 
PDMS silicone elastomer kits Sylgard 184 (SG184) and Sylgard 186 (SG186) from Dow were used and mixed according to the manufacturer's specifications.
The degassed PDMS mixtures were spin coated on the cleaned Si wafer cuts for $5~\si{\minute}$ at spin frequencies of $\omega = 6000~\mathrm{rpm}$ for SG184 and $\omega = 8000\,\mathrm{rpm}$ for SG186 to obtain uniform film thicknesses of $(7 \pm 2)~\si{\micro\meter}$ (measured by AFM) and cured on a hot plate at $75~\si{\celsius}$ for $90~\si{\minute}$. 

The shear modulus of PDMS bulk samples were determined using a frequency sweep test using a Haake-Mars-40 rheometer in plate-plate geometry with a radius of $25 \ \si{\milli\meter}$. To guarantee good mechanical contact of PDMS and shear geometry, the PDMS mixtures were cured in the shear geometry at $80~\si{\celsius}$ to obtain a shear modulus of $\Gstiff=595$~kPa for Sylgard~184 and of $\Gsoft=224$~kPa for Sylgard~186.

Small PS droplets on PDMS were prepared by dewetting. For that, PS layer with a thickness of $120~\si{\nano\meter}$ were first prepared in a glassy state by spin coating a PS-toluene solution on a freshly cleaved mica sheet. The thus prepared glassy PS layer is then floated on an ultra-pure water interface (Fisher Scientific) and picked up from there with a previously prepared PDMS coated Si substrate. The used PS, purchased from Polymer Standards Service (Mainz, Germany), has a molecular weight of $17.8~\si{\kilo\gram\per\mol}$ and a monodispersity of $M_w/M_n=1.04$. To avoid contamination, sample preparation was done in an ISO~5 clean room atmosphere. The prepared PS/PDMS samples were annealed at $T_\mathrm{a}=(120 \pm 1)~\si{\celsius}$, i.e. about $20~\si{\celsius}$ above the glass-transition temperature of PS(17.8k). During the annealing period, the initially uniform PS layer becomes liquid and transforms into droplets by dewetting from the underlying PDMS substrate. To ensure that the obtained PS droplets are in, or very close to equilibrium, the SG184 samples were typically annealed for two days and the obtained droplets seem perfectly round. Addition experiments with annealing periods of up to eight days revealed no changes after the second day. In case of SG186 samples, the dewetting is substantially slower and annealing times of 20 days were applied. However, full equilibration exceeded experimentally accessible time scales and even after more than 20 days of annealing, the history of the dewetting pathway is still visible by a slightly elliptical footage in the horizontal plane of the droplets resulting in contact angles that vary slightly between the long and the short axis of the droplet, see \Cref{AFM-SI}.

The 3D shape of the obtained equilibrium PS droplets sitting on the PDMS substrate were analyzed by atomic force microscopy (AFM, Bruker Dimension FastScan), using Olympus tips (OMCL-AC160TS-R3) in Soft Tapping Mode. AFM scanning was typically obtained at room temperature, where the PS droplets are in a glassy state, which reduces the probability of contaminating the probes. However, test experiments were also carried out at dewetting temperatures of $T=120~\si{\celsius}$, where the droplets were in the liquid state, and no difference in the droplet shape were observed.

To additionally image the buried PS-PDMS interface of the droplets, a lift-off technique was applied. To this end, a UV-curable glue layer (Norland optical adhesive, NOA~60) was poured on top of the PDMS sample embedding the glassy PS droplets. The glue is cured at a wavelength of $\lambda = 366~\si{\nano\meter}$ (Benda UV lamp) for $15~\si{\minute}$. Removing the cured glue from the PDMS sample, the PS droplets remain attached to the glue and can be lifted off from the PDMS substrate, and imaged by AFM. 
The shape of the formerly PS-PDMS and PDMS-air interfaces is aligned with the previously scanned PS-air and PDMS-air interfaces to construct a complete 3D shape; see also the schematic diagram in \Cref{AFM-SI}.

To obtain information about the chemical composition of our sample surface, the  NanoIR (Bruker) technique was used. This technique uses frequency tuned infrared laser pulses to heat the top layer of the sample that is most pronounced in case of resonance absorption. The local heating results in thermal expansion, i.e., in a mechanical response of the surface, proportional to the frequency dependent absorption coefficient. This way a relative IR spectrum between 2 and 20 micrometers can be obtained with the lateral resolution of an AFM.

\subsection{Theoretical Models}
In the following, we set up a model for the computation of stationary liquid droplets and viscoelastic substrates with moving capillary interfaces. With the goal to describe a simplified relaxation into a stationary state, we will use a sharp-interface model that features an initially flat substrate $\Omega^0_\mathrm{s}$ and a liquid droplet $\Omega^0_{\ell}$ which is the half of an ellipsoid. The viscoelastic substrate is supported by a rigid wafer $\Omega^0_\mathrm{w}$ surrounded by an ambient air domain $\Omega^0_\mathrm{a}=\mathbb{R}^3\setminus(\Omega_\mathrm{s}\cup\Omega_\mathrm{w}\cup\Omega_\ell)$ as explained in \Cref{fig:sketch}. Between different domains we define interfaces $\Gamma^0_{ij}={\Omega}^0_i\cap{\Omega}^0_j$ for $i,j\in\{{\rm a,\ell,s,w}\}$. In particular, the initial undeformed substrate $\Omega^0_\mathrm{s}$ is a reference domain of stress-free elastic material.

\definecolor{dropcol}{HTML}{0072A8}
\definecolor{subscol}{HTML}{CDC7B9}
\begin{figure*}[ht!]
    \centering
    \begin{tikzpicture}[scale=1.9,every node/.style={scale=0.8}]
        \fill [color=dropcol] (1,1,0) arc (0:90:1) -- (0,1,0) -- cycle;
        \fill [color=subscol] (0,0) rectangle (2.5,1);

        \node[gray] at (1.25,2.3) {$\Omega^0$};
        \draw[thick,->, black] (-0.03,0) -- (3,0) node[anchor=north]{$r$};
        \draw[thick,->, black] (0,-0.03) -- (0,2.5) node[anchor=west]{$z$};       
        \draw[](1,1,0) arc (0:91:1) {};
        \draw[gray,->](0.25,2.43,0) arc (-40:220:0.3 and 0.1) node[]{};
        \draw[thin,black] (-0.03,1) -- (2.5,1) {};
    
        \node[] at (1.2,0.5) {substrate\,\,$\Omega^0_\mathrm{s}$};
        \node[color=white] at (0.4,1.4) {liquid\,\,$\Omega^0_\ell$};
        \node[] at (1.6,1.6) {air\,\,$\Omega^0_\mathrm{a}$};
        \node[] at (-0.1,1) {$H$};
        \node[] at (-0.07,-0.07) {$0$};
        \node[] at (1.7,0.88)  {$\Gamma^0_{\rm as}$};
        \node[] at (0.5,0.88)  {$\Gamma^0_{\rm \ell s}$};
        \node[] at (0.95,1.65) {$\Gamma^0_{\rm a{\ell}}$};
        \node[] at (1.2,0.13)  {$\Gamma^0_{\rm sw}$};

      	\foreach \x in {0,0.1,...,2.5}
	 	{
	 		\draw[] (\x,0) -- (\x+0.15,-0.15);
	 	}

        \node[fill=white,rounded corners=2pt,inner sep=1pt] at (1.2,-0.12) {\,\,wafer \,\,$\Omega^0_\mathrm{w}$\,\,};
      
        \node at (1,1) [circle,fill,inner sep=1.2pt]{};
        \def\xs{4}
        \def\ys{-0.1}
        \fill [color=dropcol] (\xs,2+\ys) arc (90:0:1 and 0.8) -- (\xs+1,1.2+\ys) arc (0:-90:1 and 0.3) -- cycle;
        \fill [color=subscol] (\xs,0.9+\ys) arc (-90:0:1 and 0.3) -- (\xs+1,1.2+\ys) arc (-160:-90:1.6 and 0.2) -- (\xs+2.5,1.1+\ys) -- (\xs+2.5,0) -- (\xs,0)-- cycle; 
    
        \draw[] (\xs+1,1.2+\ys) arc (-160:-90:1.6 and 0.2);
        \node[gray] at (\xs+1.25,2.3) {$\Omega(t)$};
        \draw[thick,black,->](2.6,1.1) arc (140:40:0.7 and 0.2) node[midway,above]{$\bs{\chi}:[0,T]\times\Omega^0\to\mathbb{R}^3$};
        \draw[thick,->, black] (\xs+-0.03,0) -- (\xs+3,0) node[anchor=north]{${r}$};
        \draw[thick,->, black] (\xs+0,-0.03) -- (\xs+0,2.5) node[anchor=west]{${z}$};
        \draw[gray,->](\xs+0.25,2.43,0) arc (-40:220:0.3 and 0.1) node[]{};
        
        \node[] at (\xs+1.3,0.5) {${\Omega}_\mathrm{s}$};
        \node[color=white] at (\xs+0.3,1.3) {${\Omega}_\ell$};
        \node[] at (\xs+1.6,1.6) {${\Omega}_\mathrm{a}$};
        \node[] at (\xs-0.07,-0.07) {$0$};
        \node[] at (\xs+2.30,0.95+\ys) {$\Gamma_{\rm as}$};
        \node[] at (\xs+0.15,0.8+\ys) {$\Gamma_{\rm \ell s}$};
        \node[] at (\xs+0.15,2.12+\ys) {$\Gamma_{\rm a{\ell}}$};
        \node[] at (\xs+2.3,0.13) {$\Gamma_{\rm sw}$};
        \node[] at (\xs+1.09,0.95) {$\vartheta_{\rm s}$};
        \node[color=white] at (\xs+0.87,1.15) {$\vartheta_\ell$};

        \draw (\xs,2+\ys) arc (90:0:1 and 0.8) -- (\xs+1,1.2+\ys) arc (0:-90:1 and 0.3);
        \foreach \x in {0,0.1,...,2.5}
	       {
	           \draw[] (\xs+\x,0) -- (\xs+\x+0.15,-0.15);
	       }

        \node[fill=white,rounded corners=2pt,inner sep=1pt] at (\xs+1.2,-0.12) {\,\,wafer \,\,$\Omega_\mathrm{w}$\,\,};

       \node at (\xs+1,1.2+\ys) [circle,fill,inner sep=1.2pt]{};
       \draw[dashed,black,thick] (\xs+1,1.2+\ys) -- (\xs+1+0.4,1.2+\ys-0.13); 
       \draw[dashed,black,thick] (\xs+1,1.2+\ys) -- (\xs+1-0.08,1.2+\ys-0.4); 
       \draw[dashed,black,thick] (\xs+1,1.2+\ys) -- (\xs+1,1.2+\ys+0.4); 
       \draw[dotted,white,thick] (\xs+1,1.2+\ys+0.3) arc (90:340:0.3); 
       
    \end{tikzpicture}
    \caption{Sketch of axisymmetric liquid droplet (blue shading) on a viscoelastic substrate (gray  shading) with contact line (black dot) and surrounding air phase (white) on a rigid wafer (stripe pattern) and capillary interfaces (black lines). The left side shows the reference domain (Lagrangian)
    $\Omega^0$ and the right side shows the deformed configuration (Eulerian) $\Omega$ and all referential and deformed interfaces 
    $\Gamma_{ij}(t)=\bs{\chi}(t,\Gamma^0_{ij})$ for $i,j\in\{\rm a,\ell,s,w\}$. We denote the solid contact angle between $\Gamma_{\rm as}$ and 
    $\Gamma_{\rm \ell s}$ by $\vartheta_{\rm s}$ and the liquid contact angle between $\Gamma_{\rm \ell s}$ and $\Gamma_{\rm a\ell }$ by $\vartheta_\ell$. 
    The substrate is initially flat  $\Omega^0_\mathrm{s}=\{\bs{x}\in\mathbb{R}^3:0\le z\le 1\}$ and the liquid droplet is the half of an ellipsoid 
    $\Omega^0_\ell=\{\bs{x}\in \mathbb{R}^3:\nicefrac{(x^2+y^2)}{r_x^2} + \nicefrac{(z-H)^2}{r_z^2}\le 1 \text{ and }z\ge H\}$   
    with $\bs{x}=(x,y,z)$. 
    The  substrate is supported by a rigid wafer $\Omega^0_\mathrm{w}=\{\bs{x}\in\mathbb{R}^3:z\le 0\}$ surrounded by an ambient air domain $\Omega^0_\mathrm{a}=\mathbb{R}^3\setminus(\Omega_\mathrm{s}\cup\Omega_\mathrm{w}\cup\Omega_\ell)$.
    }
    \label{fig:sketch}
\end{figure*}
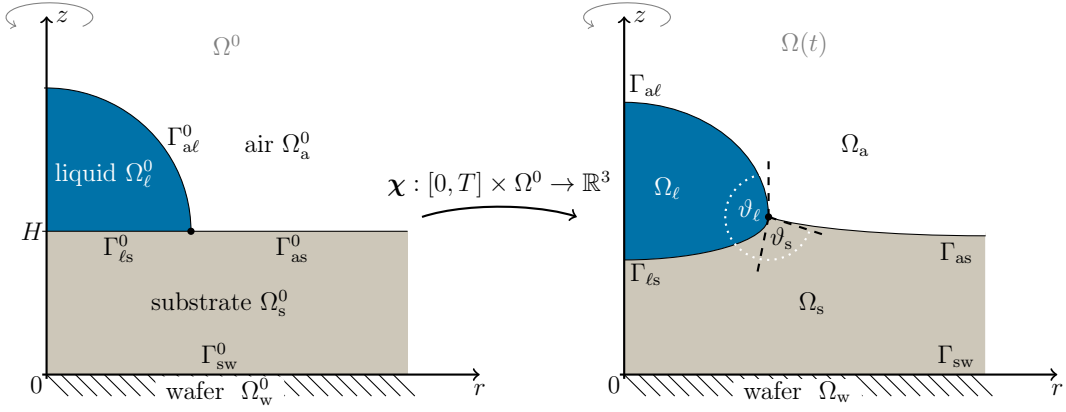

On the (axisymmetric computational) domain $\Omega^0=\Omega^0_{\rm s}\cup\Omega^0_\ell$ we define the deformation $\bs{\chi}:[0,T]\times\Omega^0\to\mathbb{R}^3$, which maps the initial solid and the liquid domain to the time-dependent domains 
$\Omega_i(t)=\bs{\chi}(t,\Omega^0_{i})$ and correspondingly the interfaces to the time-dependent interfaces $\Gamma_{ij}(t)=\bs{\chi}(t,\Gamma^0_{ij})$ with the goal to determine the stationary shapes as $t\to\infty$. Throughout the paper we denote the deformation gradient by $\bs{F}:=\nabla\bs{\chi}$.

With the given geometrical definitions, the cornerstone of the sharp-interface model is the free energy $\mathcal{F}$, which for a given deformation $\bs{\chi}$ measures the  elastic energy and the surface energy. It is defined as

\begin{align}
\nonumber\mathcal{F}(\bs{\chi}) = & \int_{\Omega^0_{\rm s}}\!W_{\text{elast}}(\boldsymbol{F})\,{\rm d}\bs{x} + \int_{\Omega^0}\!\tfrac{K}{2}(\det\bs{F}-1)^2\mathrm{d}\bs{x}\\  
\label{equ:sharp_lagrange_energy}& + \sum_{ij\in\{{\rm \ell s,\,as,\,a\ell }\}}\int_{\Gamma^0_{ij}} W^{\Gamma}_{ij}(\bs{F},\bs{\nu})\,{\rm d}\bs{a}\,,
\end{align}
where $\bs{\nu}$ is a normal vector field on respective interface $\Gamma_{ij}$ and $K\gg G$ is the bulk modulus of the nearly incompressible materials. We assume that the elastic energy density $W_{\text{elast}}$ is of neo-Hookean type and the surface/interface energy density $W^{\Gamma}_{ij}$ measures the deformed interface length multiplied with the surface tension coefficient, which gives rise to 
\begin{subequations}
\begin{align} 
\label{equ:F_elast}&W_{\text{elast}}(\boldsymbol{F}) 
     = \frac{G}{2}{\rm tr}(\boldsymbol{F}^T\boldsymbol{F}-\bs{I}),\\
\label{equ:F_surf} &W^{\Gamma}_{ij}(\bs{F},\bs{\nu}) = \gamma_{ij}|\mathrm{cof}(\boldsymbol{F})\cdot\boldsymbol{\nu}|.
\end{align}
\end{subequations}

Here, $G$ is the shear modulus of the substrate, $\bs{I}\in\mathbb{R}^{3\times 3}$ the identity matrix and we denote $\mathrm{cof}(\bs{F})=\det(\bs{F})\bs{F}^{-1}$ the cofactor matrix of $\bs{F}\in\mathbb{R}^{3\times 3}$. Upon non-dimensionalization of the free energy, the elastocapillary length \eqref{eqn:elastocap} 
appears as a key parameter of the system. We summarize the main physical assumptions that are used in this model:
\begin{enumerate}[i.)]
\item At the substrate-liquid interface  we assume a no-slip boundary condition, i.e. $\bs{\chi}(t,\bs{x})$ is continuous at $\bs{x}\in\gammals^0$. On the substrate-waver interface, we have a no-slip condition $\bs{\chi}(t,\bs{x})=\bs{x}$ for $\bs{x}\in\Gamma^0_{\rm sw}$.
\item The viscoelastic substrate is hyperelastic, i.e. the deformation in equilibrium is determined by the minimization of an elastic energy $W_{\rm elast}$ that depends on the deformation gradient $\bF$. In particular, we do not consider inelastic deformations.

\item The substrate and the liquid are both nearly incompressible $K\gg 1$.

\item The interface tensions $\gamma_{ij}$ are assumed constant, the liquid is Newtonian, and  diffusive effects in the substrate are neglected. The adoption of the specific $W^{\Gamma}_{ij}$ in \eqref{equ:F_surf} results in a spatially uniform Eulerian surface energy density, thereby excluding the Shuttleworth effect. Nevertheless, when deformation gradient dependence is considered and adequate material data is available, exploring more complex forms of the surface energy as described in \eqref{equ:F_surf} becomes feasible and relevant, e.g. cf. \cite{andreotti2016soft}.

\end{enumerate}

The main goal of our theoretical considerations is to compute equilibrium states, which minimize the free energy \eqref{equ:sharp_lagrange_energy}. Therefore, we use a simple but robust transient model that solves for $\bs{\chi}(t)$ as $t\to\infty$. 
 
Motivated by the energetical structure of the model with the free energy \eqref{equ:sharp_lagrange_energy}, we use a dynamical model for the evolution that satisfies the weak form
\begin{align} 
\label{eqn:extgrad_si}  
    s(\partial_t \bs{\chi}, \boldsymbol{v})
    :=
    \sum_{i\in\{s,\,\ell\}}\int_{\Omega_i} \mu_i\,
\nabla\partial_t\bs{\chi}:\nabla\boldsymbol{v}\,{\rm d}\bs{x}=
    -\langle {\rm D}\mathcal{F}(\bs{\chi}),\boldsymbol{v}\rangle\,,
\end{align}
for all suitable test velocity vector fields $\boldsymbol{v}$ with given initial values $\bs{\chi}(t=0)$. 
To facilitate convergence to a stationary state, we use a Kelvin-Voigt-type dissipation with material-dependent viscosities $\mu_i \in \mathbb{R}$, which we set equal to $\mu_i = \mu$.
By testing \Cref{eqn:extgrad_si} with $\bs{v}=\partial_t \bs{\chi}$ we obtain thermodynamic consistency
\begin{align}
\frac{\rm d}{{\rm d}t}\mathcal{F}\bigl(\bs{\chi}(t)\bigr)
= - s(\partial_t\bs{\chi},\partial_t\bs{\chi})\le 0.
\end{align}
For a dynamic model moving contact lines that are not pinned, the assumption i.) presents a serious restriction. 
We overcome this restriction for axisymmetric minimizers by choosing the initial ellipsoid shape, i.e., $r_x$ and $r_z$, such that for a given drop volume the stationary free energy is minimized, thereby satisfying the third \emph{no-pinning} boundary condition discussed in \cite{snoeijer2018paradox,pandey2020singular}.
In \Cref{sec:axisymmetric_discrete} we provide additional details on the spatial and temporal discretization of \eqref{eqn:extgrad_si}, implemented using the FEniCS finite element library \cite{logg2012automated}.

\section{Stationary Droplet Shapes}
\begin{figure}[b!]
\includegraphics[width=0.495\textwidth,trim={0.6cm 3.0cm 1.7cm 3.5cm},clip]{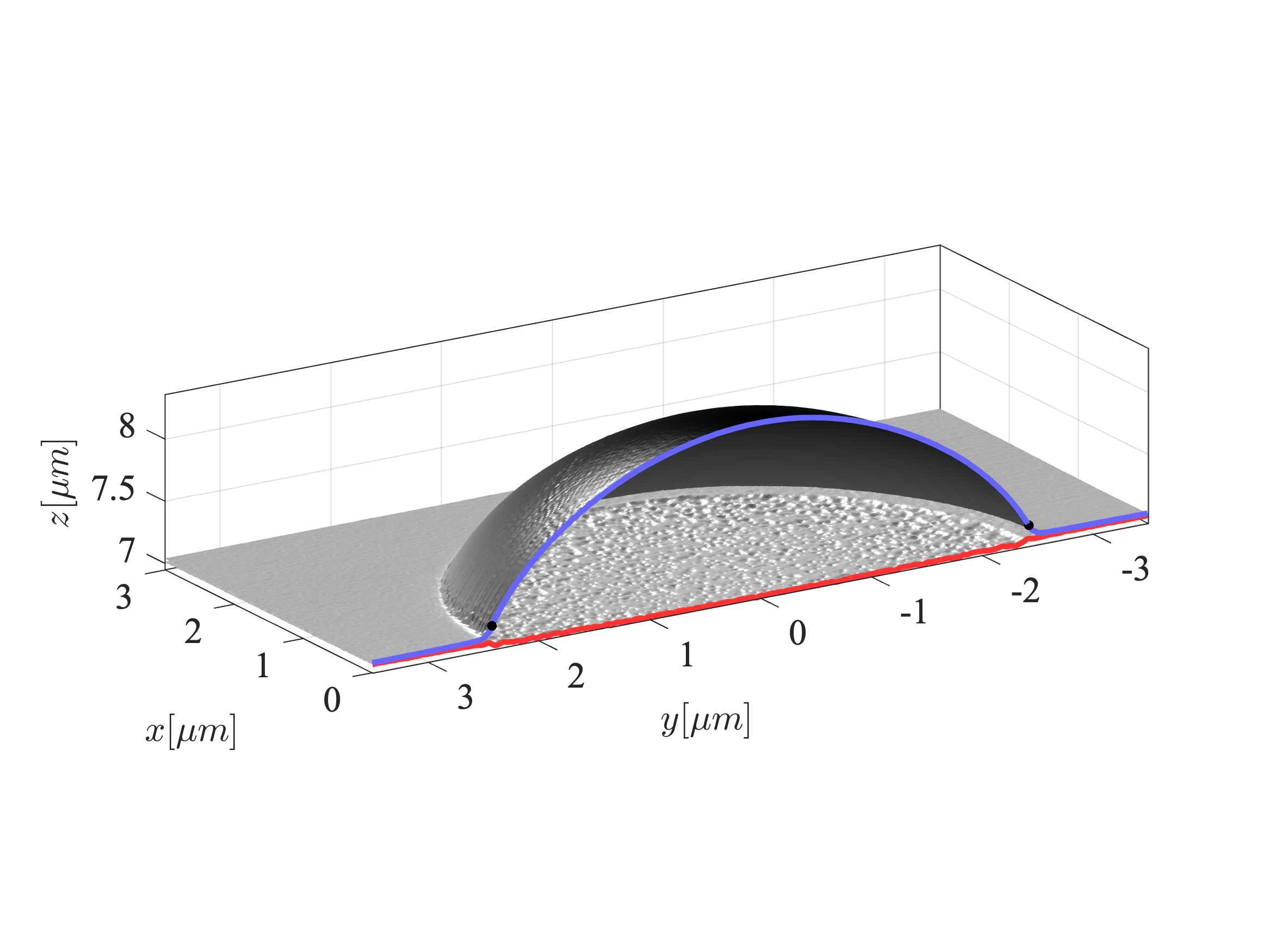}\hfill
\includegraphics[width=0.495\textwidth,trim={0.6cm 3.0cm 1.7cm 3.5cm},clip]{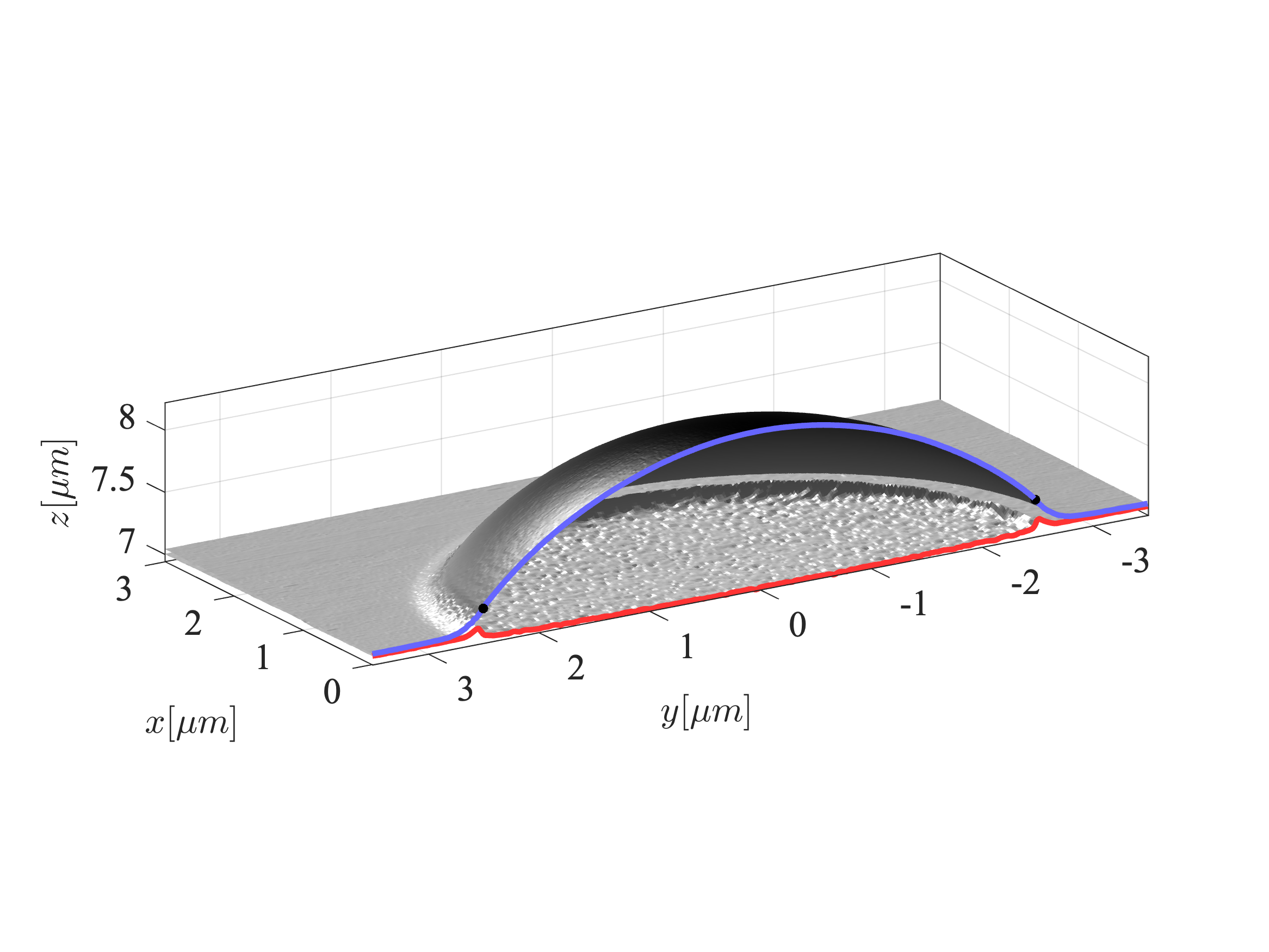}
    
\caption{3D cross sections of top and bottom AFM scans of PS droplets with radius $R\approx 2.5\,\si{\micro\meter}$ on SG184 (left) and SG186 (right); substrate height is $H\approx 7\,\si{\micro\meter}$. Blue and red lines show the central top and bottom AFM scan lines, whereas the black dots denote the positions of the contact line identified by the inflection points of the top scan line.}
\label{fig:view3d}
\end{figure}

%
%
Three-dimensional representations of experimentally obtained equilibrium droplet shapes on SG184 and SG186 substrates are shown in \Cref{fig:view3d} for the largest sets of droplet radii.
These configurations are characterized by $R \approx 2.5\,\si{\micro\meter}$, $\lambdacap = \gammaas / G \sim 30\,\si{\nano\meter}$ for SG184, and $\lambdacap = \gammaas / G \sim 80\,\si{\nano\meter}$ for SG186, corresponding to the regime of a {\em moderately soft substrate}, where $a \ll \lambdacap \ll R$ and the substrate thickness $H=7\,\si{\micro\meter}$ is larger than the droplet radius but not in the semi-infinite substrate thickness regime.

We note that within a dewetting time of several hours at $T_\mathrm{a}=(120 \pm 1)\,\si{\celsius}$, droplets on SG184 always reached a rotationally symmetric shape, while the equilibration of droplets on SG186 was not fully completed and the droplet bases remained slightly elliptical even after several days. The corresponding largest and smallest radii for PS droplets on SG184 vary by $3\,\%$, i.e. within the accuracy of the AFM, while the drop radii on SG186 vary by about $13\,\%$, see the 3D images and cross sections in \Cref{AFM-SI}.
For the large droplets with $R\gg\lambdacap$ shown in  \Cref{fig:view3d}, a spherical arc fit can be used to determine effective Young angles, see also \Cref{AFM-SI}.

The PS-PDMS interface of the droplet with the elastic substrate is deformed downwards by the Laplace-pressure inside the droplet. This deformation is relatively small because the elasto-capillary length $\lambdacap$ is small compared to the drop radius $R$.  
We also note that all AFM scans of the PS-PDMS interface show some roughness in comparison to the PS-air and PDMS-air interface, which may indicate low PS-PDMS interfacial tension but could also be due to aggregation of filler particles. 

The experimental AFM profiles of the top surface transition smoothly from the PS-air interface to the PDMS-air interface and, in particular, do not show any kink at the TPCL.
This however, does not allow to directly determine the position of the TPCL.
Thus, the TPCL location is determined indirectly by the turning point of the top AFM scan, cf. also \Cref{AFM-SI}.
The corresponding figure there shows that the turning point yields a clearly distinguished location for the position of the TPCL. The determined TPCL is indicated by the black dot on top of the blue scan line in \Cref{fig:view3d}.
%
%
The TPCL is pulled upwards with respect to the undeformed PDMS substrate to accommodate the forces generated by the surface tensions $\gamma_{ij}$ for $i,j\in \{a,s,\ell\}$ for the interfaces between the air ($a$), substrate ($s$) and liquid ($\ell$) phases. The balance of interfacial forces leads to a Neumann triangle, that generates the characteristic elastic ridge below the TPCL \cite{style2012static}. 
Furthermore, note that the shown top contours align with the bottom contour of the soft substrate at a slight distance from the droplet and the TPCL. However, in the immediate vicinity of the TPCL shown in \Cref{fig:view3d}, the upper and lower AFM profiles do not align, creating the impression that a small volume of PDMS is missing. This missing alignment near the TPCL is present in all experiments and is enhanced for the softer SG186 substrate.


\begin{table*}
\centering
\renewcommand{\arraystretch}{1.5}
  \begin{tabular}{r|c|ccc}
    \hline\hline
     & surface tension $[\si{\milli\newton\per\meter}]$ &  \multicolumn{3}{c}{surface tension $[\si{\milli\newton\per\meter}]$} \\
    interface & literature & SG184 & & SG186 \\ 
    \hline
PDMS-air, $\gammaas$ & $15 \pm 1$ \,\cite{bhatia1985measurement,roe1968surface}  &$15\,(\pm 1)$ & 
&$15\,(\pm 1)$\\
PS-air, $\gammaal$ / $\gammaal^c$ & $31\pm 1$ 
\, \cite{wu1970surface,equilibrium,dee1992molecular} & $19.2 \pm 0.1\,(\pm 1.3)$ & 
& $18.8/17.8\,(\pm 1.2)$ \\
PS-PDMS, $\gammals$ & $0-10$ \, \cite{nose1997temperature,nose1996temperature}&$4.2 \pm 0.1 \,(\pm 0.3)$ & & $3.8/2.8\,(\pm 0.2)$\\
    \hline\hline
  \end{tabular}
\caption{Interfacial tensions at $T_\mathrm{a}=120\si{\celsius}$ from literature (left column) and from hybrid construction  \eqref{eq:hybrid} for large cloaked droplets (right column) 
for SG184 and SG 186. Errors in the right column are based on Young angle along the shorter/longer axis and the errors in parenthesis {additionally} represent the uncertainty of the reference value, $\gammaas$.}
\label{tab:params_surf_litx}
\end{table*}

For equilibrium droplet shapes, the relevant system parameters are the interfacial tensions of PS-PDMS, PS-air, and PDMS-air and the shear modulus $G$ of PDMS.
In \Cref{tab:params_surf_litx} we list the literature values for the corresponding interfacial tensions, which are not compatible with an equilibrium Neumann construction since the stability condition $\gammaas+\gammals\ge\gammaal$ is violated, cf. \Cref{fig:sketch}.
This means that, based on these literature values, stable PS droplet configurations on PDMS that adhere to the Neumann construction are impossible, which is in clear contrast to our experimental observations shown in \Cref{fig:view3d}. 

\begin{figure}[t]
\includegraphics[height=0.24\textwidth,clip]{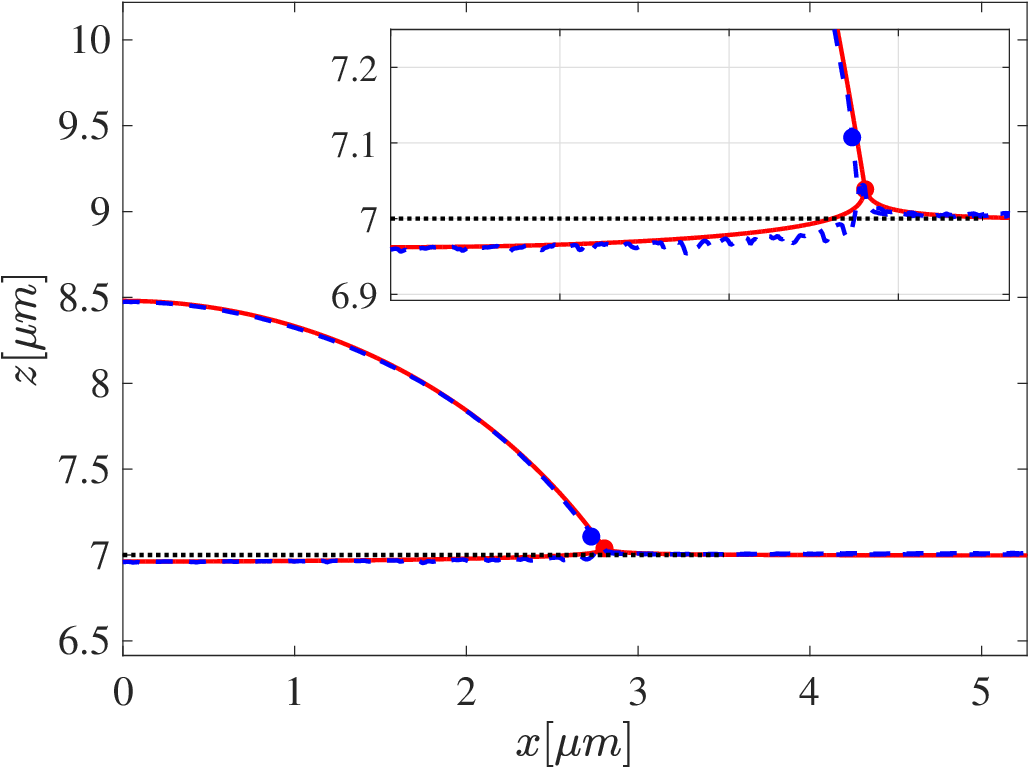}\hfill%
\includegraphics[height=0.24\textwidth,clip]{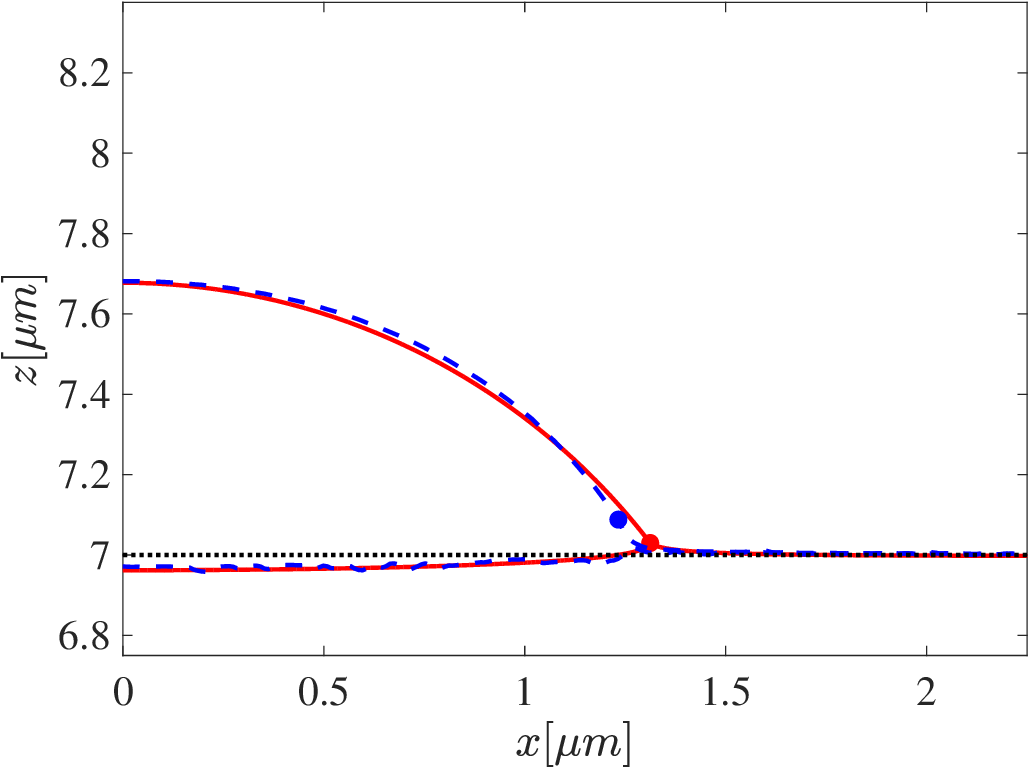}\hfill%
\includegraphics[height=0.24\textwidth,clip]{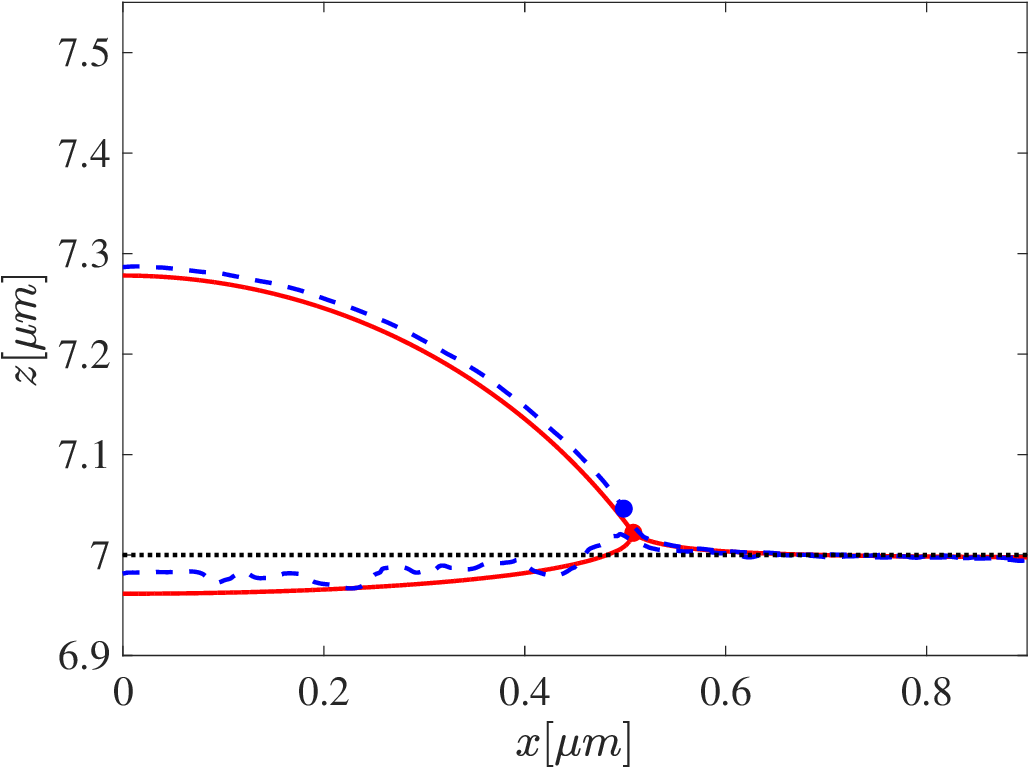}%

\includegraphics[height=0.24\textwidth,clip]{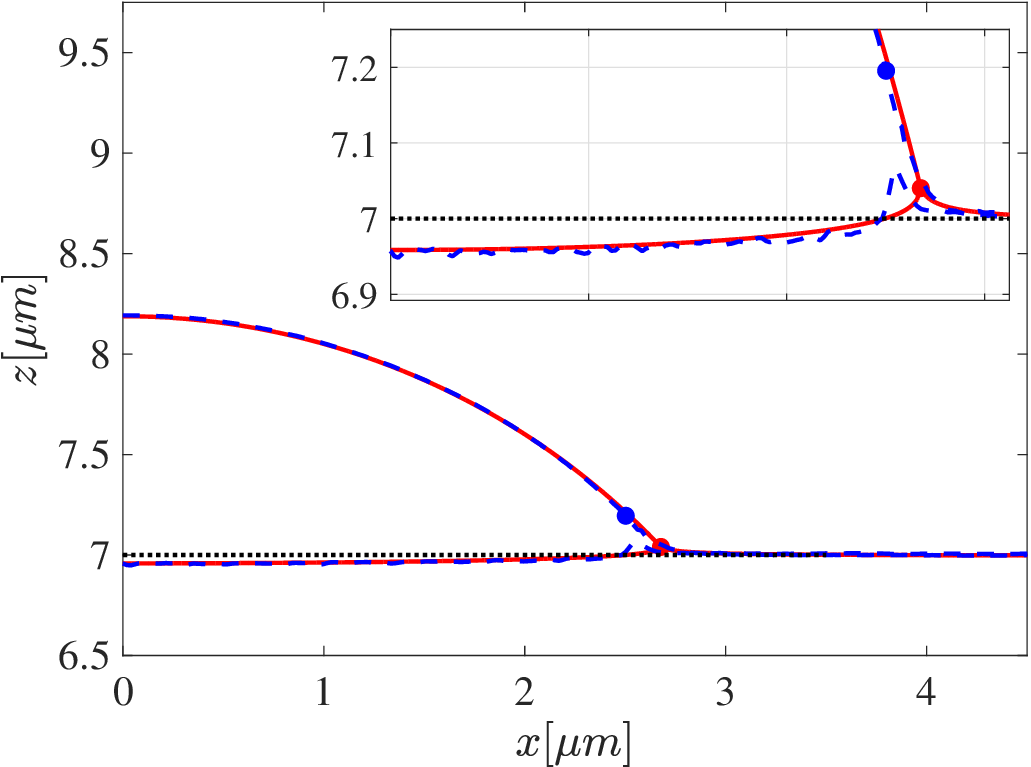}\hfill%
\includegraphics[height=0.24\textwidth,clip]{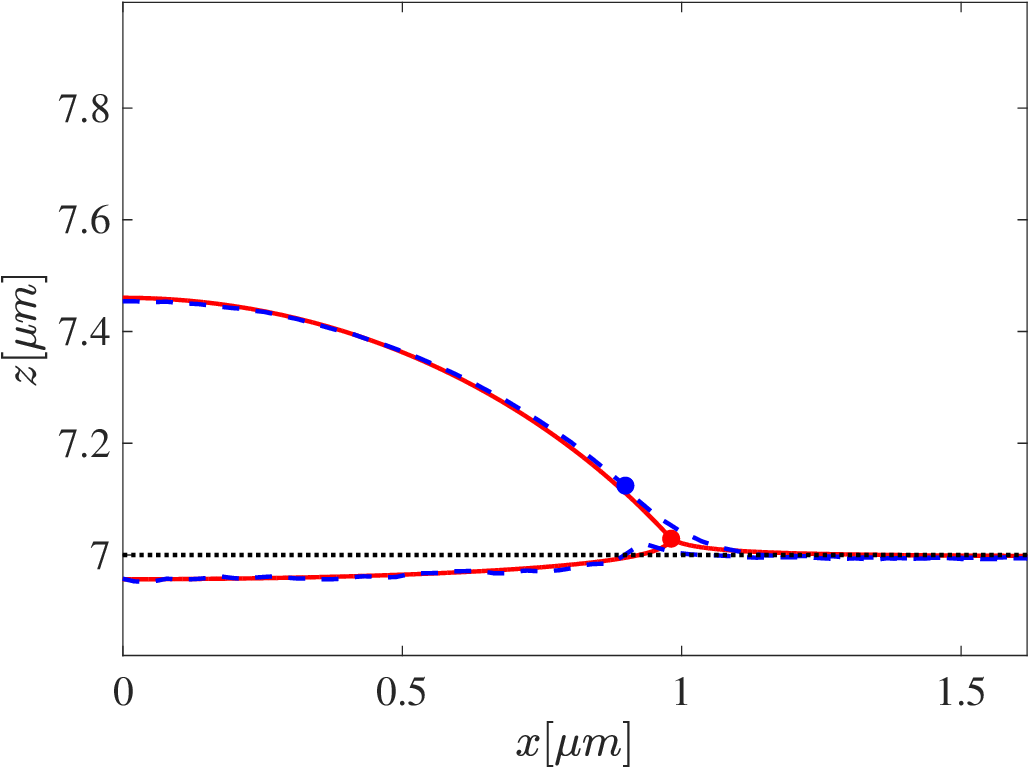}\hfill%
\includegraphics[height=0.24\textwidth,clip]{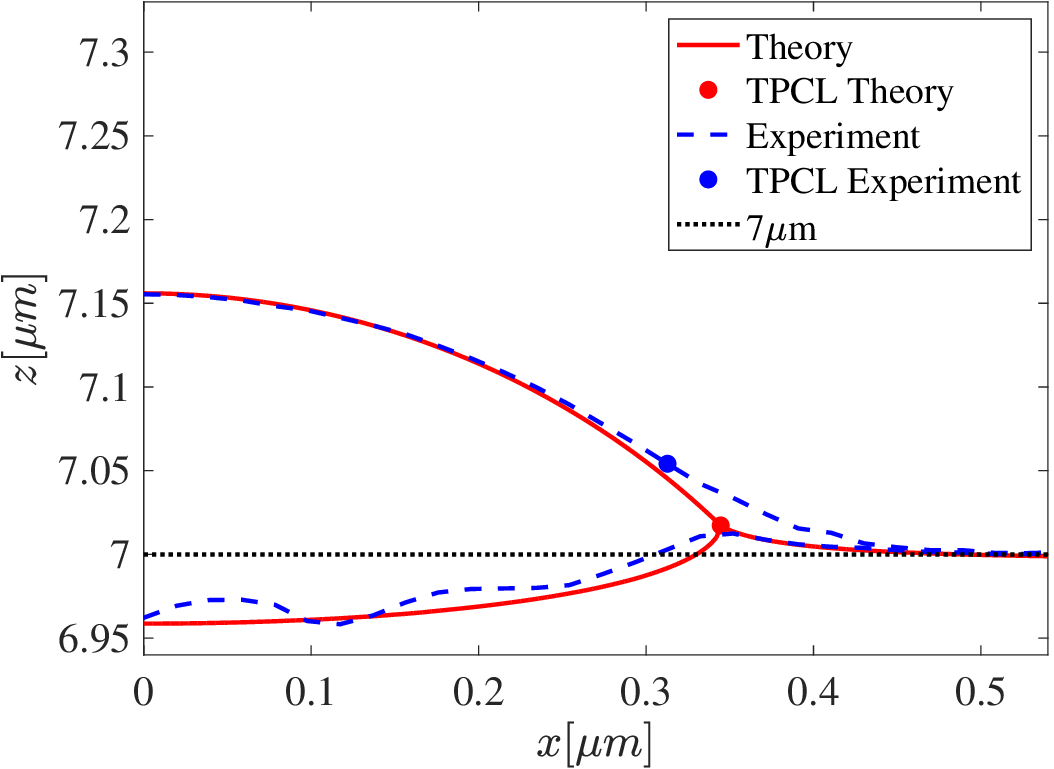}

\caption{Comparison of experimental AFM cross section (blue dashed line) for Sylgard 184 (top) and Sylgard 186 (bottom)  and theoretical prediction (red line) for different droplet radii decreasing from left to right. The position of the TPCL is shown using a dot and the inset shows a close-up of the liquid-substrate interface.
The horizontal dotted line shows the undeformed PS-PDMS/undeformed PDMS-air interface assumed  at exactly  $z=7\,\si{\micro\meter}$.}
\label{fig:comp_exp_theo}
\end{figure}

Using NanoIR (Bruker), we observe a layer of liquid PDMS along the PS-air interface, a phenomenon known as cloaking, see \Cref{Chemical-Composition}. 
The observed strength of the PDMS signal suggests that the thickness of the cloaking layer is thick enough to lower the surface tension of the PS-air interface throughout the droplet-air interface from $(31\pm1)~\si{\milli\newton\per\meter}$ \cite{wu1970surface,equilibrium,dee1992molecular} to a fully cloaked PS-air interface with
\begin{subequations}
\label{eq:hybrid}
\begin{align}
\label{eqn:hybrid_cloak}
\gammaal^{c} = \gammals+\gammaas,
\end{align}
equivalent to a degenerate Neumann construction with $\vartheta_\mathrm{s}=0$ and $\vartheta_\ell=180\si{\degree}$, see \Cref{fig:sketch}.  
Therefore, the relation \eqref{eqn:hybrid_cloak} corresponds to the experimentally observed 
smooth transition from the PDMS/air interface to the cloaked PS/air interface that can be seen in \Cref{fig:view3d}. 
However, since the value of $\gammaal^{c}$ directly depends on $\gammals$, which in turn varies with the chain length and temperature of PS and PDMS, cf. \Cref{tab:params_surf_litx} and \cite{nose1997temperature,nose1996temperature}, we can only provide a relatively broad range for the surface tension at typical dewetting temperatures of $T_\mathrm{a} = (120 \pm 1),\si{\celsius}$.
The PS-PDMS interfacial tension, $\gammals$, for our system is determined using a hybrid construction that applies the effective Young contact angle $\theta$ to a horizontal substrate. This construction is well-defined only for $R \gg \lambdacap$, i.e., for large droplets where the deformation of the PDMS substrate is relatively small compared to $\lambdacap$, see e.g. \cite{style2012static}. Using Young's equation,
Using a hybrid construction that applies the effective Young contact angle $\theta$ to a horizontal substrate, the PS-PDMS interfacial tension, $\gammals$, can be determined specifically for our system.
This construction is well-defined only for $R \gg \lambdacap$, i.e., for large droplets where the deformation of the PDMS substrate is relatively small compared to $\lambdacap$, see e.g. \cite{style2012static}. Using Young's equation,
\begin{align}\label{equ:young}
\gammals = \gammaas - \gammaal^c\cos(\theta),
\end{align}
\end{subequations}
we calculate $\gammals$ by reconstructing the Young angle $\theta$ through the fitting of spherical arcs to radial cross-sections of experimental droplet shapes. For SG184, this yields $\theta = (55.9 \pm 0.3) \si{\degree}$, with the error estimated by comparing spherical arcs fitted along the longest and shortest radial cross-section of the measured droplet profiles.
We note that for PS droplets on SG186, we observe stronger deviations from an axisymmetric droplet shape, so that the contact angle for the largest droplets vary between $\theta=47.0\si{\degree}$ (long axis) and $\theta=52.7\si{\degree}$ (short axis), see \Cref{AFM-SI}. 
The resulting surface tensions obtained by this hybrid construction, using \eqref{eq:hybrid}, are given in the right column of \Cref{tab:params_surf_litx}.

To quantitatively match numerical and experimental droplet shapes, the three-di\-men\-sion\-al volume of each experimental droplet was computed under the assumption of axisymmetry of the cross-section along a given scan direction, resulting in the volumes provided in the table in \Cref{sec:axisymmetric_discrete}.
These volumes were converted into radii $r_x, r_z$ used for initial data described in \Cref{fig:sketch}. Accordingly, for a given shear modulus $G$ and interfacial tensions $\gamma_{ij}$, stationary solutions were computed solving \eqref{eqn:extgrad_si} leading the energy minimizers of \eqref{equ:sharp_lagrange_energy}.
For the large droplets $R=(2-3)\,\si{\micro\meter}$ with  $\lambdacap\ll R$, the PS-PDMS interface reveals a small but in the close up clearly visible depression below the droplet, \Cref{fig:view3d}. Cross sections of experimentally measured depressions of the largest droplets are used to fit the shear modulus $G$ of the PDMS substrate that provides the best match between numerical and experimental data. For SG184, the thus obtained shear modulus
is $50\%$ lower than the experimentally measured one, $\Gstiff=298\,\si{\kilo\pascal}=0.5 \, \Gstiff^\text{exp}$. 
For SG186, we find that the fitted value agrees well with the experimentally determined value, $\Gsoft=\Gsoft^\text{exp}=224\,\si{\kilo\pascal}$. 
These constant shear moduli for SG184 and SG186 from large droplets are then used to compute stationary states of smaller droplets without any further adjustments.

%

With the determined experimental droplet volumes provided in \Cref{sec:axisymmetric_discrete}, the hybrid construction to determine surface tensions and the adjusted shear moduli, we have derived theoretical stationary droplet configurations that can be compared with the corresponding experimental results as shown in \Cref{fig:comp_exp_theo}. 
Droplet shapes computed for different volumes on SG184 and SG186 substrates are shown by red solid lines in \Cref{fig:comp_exp_theo} 
together with the corresponding experimental droplet shapes (blue dashed lines). Corresponding experimental and theoretical TPCL positions are shown with the same color coding using solid dots. For the largest droplet radii on SG184 and SG186,  the inset highlights that matching indentation of the PS-PDMS interface.

All shown configurations on SG184 and SG186 show  excellent overall agreement for the determined 
interface shapes for the PS-air, PS-PDMS and PDMS-air interface. Within the limits given by the interfacial roughness, the indentation of the PS-PDMS is well captured and the constant curvature of the PS-air interface perfectly matches the theoretical prediction of the axisymmetric theoretical profiles. 
While large droplets $R>1\,\si{\micro\meter}$, shown in the left and middle panel of \Cref{fig:comp_exp_theo}, look qualitatively similar, the smaller droplets with $R<0.5\,\si{\micro\meter}$ in the right panel show the transition from moderately soft substrates towards the soft limit in terms of stronger relative deformation of the PS-PDMS interface. 
Droplets with $R \ll \lambda_{\text{cap}}$, much smaller than those probed experimentally,  assume equilibrium shapes similar to the liquid lenses on liquid substrates \cite{equilibrium}, e.g. see right panel of \Cref{fig:theoretical_droplets_intro}.

Finally, we note that the softer SG186 generally shows larger deformations of the PS-PDMS interface for similarly-sized droplets. One general difference between experimental and theoretical interface shapes is that, although the theoretical prediction employs a degenerate  Neumann triangle construction \eqref{eqn:hybrid_cloak}, the transition from the PS-air to the PDMS-air interface does not appear as smooth as in the experiments on the scale of the droplet radius. 

In the following, we will use the general good agreement of the global experimental and theoretical droplet shapes and focus on the discrepancies near the TPCL.
%
While the trends and interface shapes for different droplet sizes agree well with the experimental results for all substrates, we observe a systematic deviation in the immediate vicinity of the TPCL. This trend enhances for decreasing droplet size and for softer substrates, see \Cref{fig:comp_exp_theo}.
The exact shape of the PS-PDMS interface near the contact line appears difficult to predict due to the missing alignment of the top and bottom AFM scan at the TPCL, see \Cref{fig:view3d} and  \Cref{fig:comp_exp_theo}.
However, already from the PS-air interface, it is clear that the experimentally measured elevation of the TPCL is raised about three times higher above the unperturbed PDMS surface than in the theoretical prediction; see dots in \Cref{fig:comp_exp_theo}. In addition, the position of the TPCL in the experimental drop profiles is not only shifted upwards compared to the numerical drop profiles but also inwards, towards the center of the drop. Both the relative magnitude of the upward deformation and the inward shift of the TPCL, i.e., the deviation between numerical and experimental results shown in \Cref{fig:comp_exp_theo}, increase for smaller droplets and for softer substrates with 
droplet sizes closer to the elastocapillary length.

To investigate these discrepancies, we first varied the substrate thickness in the range $H=5-10\,\si{\micro\meter}$, see \Cref{sec:dependence_on_H},  and found only a small impact on the droplet shape and indentation and a negligible impact on the droplet shape near the TPCL, as long as $\lambdacap<H$ and $R < H$, which is satisfied for all our experimental results.
Computed PS droplet shapes on SG186 for a drastically reduced shear modulus of $\tfrac{1}{16} \Gsoft^\text{exp}$ match the elevation of the three-phase contact line, see \Cref{sec:dependence_G}. 
However, then the global drop shape fails to describe the experimental profiles, indicating that the overall substrate is too soft. 

A similar theoretical test, with varying surface tensions to account for different solid opening angles $\vartheta_s$ in \Cref{sec:theta_S}, also failed to resolve the observed discrepancies between theory and experiment at the TPCL. Verifying the actual solid opening angles $\vartheta_s$ experimentally would require a much higher spatial resolution on the molecular scale, around $a \sim 10^{-9},\si{\meter}$. This aligns with theoretical predictions, which suggest that an adaptive meshing strategy in numerical computations is necessary to capture the singular nature of the capillary ridge at $\vartheta_s = 0$ \cite{pandey2020singular}.

Thus, we can conclude that, under the aforementioned physical modeling assumptions, the theoretical model has no remaining parameters to reconcile the differences near the TPCL between experimental and numerical drop shapes while maintaining the good global agreement of the predicted droplet shapes.

\section{Discussion and Outlook}
As global parameters are unable to account for the observed mismatch between the theoretical model and the experimental results near the TPCL, localized phenomena caused by an increased elastocapillary length, $\lambdacap$ are expected to play a role. Such an increase could arise from a local reduction in shear modulus  $G$ or a local  enhancement of surface tension $\gamma_{ij}$ caused by different complex mechanisms that we discuss in the following.

A locally higher surface stress could be attributed to the Shuttleworth effect \cite{Shuttleworth-1st}, which describes a strain-dependent surface stress in (soft) solids. However, the existence and significance of the Shuttleworth effect for PDMS substrates remain a topic of debate in the literature. While the majority of studies report a positive Shuttleworth effect, i.e., a stiffening of stretched interfaces both at the PDMS-air and PDMS-PS interfaces \cite{style2013universal,Xu2017,Xu2018SoftMatter,snoeijer2018paradox,Bain2021,Heyden2021,Zhao2022}, which was even quantified to increase with a slope of about $1\,\si{\milli\newton\per\meter}$ for each percent of strain \cite{Xu2017,Xu2018SoftMatter}, others suggest an asymmetric Shuttleworth effect \cite{Park2014} or no observable Shuttleworth effect \cite{Schulman2018}. The latter argue that such an effect is unlikely in polymers with a reduced crosslinking density at the interface \cite{Schulman2018}, a scenario expected for our PDMS surfaces. However, to the best of our knowledge, there is no evidence in the literature supporting the existence of a negative Shuttleworth effect for PDMS substrates. 

The overall stretching of the PDMS in our case is less than $1\,\%$ (vertical height of the TPCL $< 100,\si{\nano\meter}$ at a PDMS thickness of $\sim 7\,\si{\micro\meter}$) and a Shuttleworth effect based on a rather global (pre-) strain would account for an increase of the surface strain of less than 1\,\si{\milli\newton\per\meter} \cite{Xu2017,Xu2018SoftMatter} and can therefore be ignored.
In contrast, the local surface elongation is higher and in the order $10-100\%$, but limited to a distance of about $<30\,\si{\nano\meter}$ around the TPCL, see  \Cref{sec:local_shear_TPCL}. So despite the expected 
Shuttleworth effect on a very local scale, its lateral extension seems way too small to account for the observed discrepancies {near the TPCL}. However, if the above reasoning were incorrect and a positive Shuttleworth effect were incorporated into the theory, the additional surface stiffening at the PS-PDMS and PDMS-air interfaces would shift the position of the computed TPCL even further downward, thereby increasing the observed vertical discrepancy between the numerical and experimental results. 
However, it has been shown in \cite{Henkel_Shuttleworth} that an asymmetric Shuttleworth effect can significantly alter the contour of the PDMS surface and could hence help to match the PDMS contour. But it was also shown in \cite{Henkel_Shuttleworth} that an asymmetric Shuttleworth effect has only minimal impact on the horizontal position of the three-phase contact line and thus fails to explain the discrepancy in this respect.
In summary, assuming an asymmetric and negative Shuttleworth effect, one could possibly explain the observed contour of the PDMS surface, but not the position of the TPCL. And as no evidence in literature can be found for a negative Shuttleworth effect for these type of substrates and only a hint towards a possible asymmetric Shuttleworth effect \cite{Park2014},  which can also not fully explain the findings, it seems very unlikely that including the Shuttleworth effect in the theoretical model would explain or reduce the differences to the experiment. 

The observed difference could be attributed to viscoelastic effects in the crosslinked polymer network, such as stress relaxation through reversible or irreversible crosslink reformation. While a clear distinction between viscoelastic and poroelastic effects is challenging, stick-slip dynamics and hysteresis are often linked to viscoelastic behavior in the literature \cite{Cy52-276_G_Karpitschka,andreotti2020statics}. Specifically, these effects are frequently associated with rate-dependent dissipation, where the rheology is characterized by a power-law model \cite{Cy52-276_G_Karpitschka} of the form $\mu = G(1 + (i\tau\omega)^n)$, with $n = 0.55$. This dissipation mechanism is expected to vanish in the stationary limit as $\omega \to 0$. However, the persistence of deviations in axisymmetric droplet shapes, as found experimentally for SG186 substrates even after long experimental times, hints at a possible viscoelastic stick-slip mechanism, potentially caused by either bulk viscoelasticity or viscoelastic adhesion at the interface.

The global reduction of the shear modulus can be rationalized by a nonuniform crosslinking density in PDMS films. For our system, the bulk volume fractions of uncrosslinked polymer are $4-5\%$ for SG184 [1:10] \cite{Wong2020,UncrosslinkedExtracting-Uncrosslinked-SG184} and $6-7\%$ for SG186 [1:10] \cite{james1974cross}. In particular, a reduced crosslinking density near the PDMS-air interface might be expected \cite{Free-chains-interface, Free-Chains-Not-Ideal-Interface}, explaining the reduced shear modulus for SG184 obtained from fitting the numerical data compared to bulk rheology measurements. From this perspective, it is rather surprising that the bulk value for $\Gsoft$ can be used without further correction; however, this might result from the not fully equilibrated PS drop shapes on SG186.

What is also discussed in the context of polydimethylsiloxane substrates is poroelasticity to account for certain dynamic or static drop features \cite{style2013universal,Zhao2022,PhaseSeparationKJensen,PhaseSeparationJPham,Hourlier-Fargette2017,Wong2020,Zhao2018,Liu2016,Xu2020,hauer2023phase}. 
Poroelastic effects can locally soften and swell the substrate, i.e. the shear modulus can depend on the fraction of uncrosslinked molecules $c$ and concentration is coupled to the local volume through a contribution that encodes the swelling of soft gels. Corresponding theoretical models describe how elastic deformation and diffusion are coupled through mechanochemical forces, e.g. \cite{Zhao2018}. This coupling generates gradients in the chemical potential that drive the diffusion of $c$, leading to a slightly higher concentration of solvent or polymer molecules near the TPCL in weakly coupled systems. In contrast, a strong elasto-chemo-capillary coupling can even lead to localized phase separation or demixing in this region. 
Besides altered elastic properties, poroeleasticity adds a time scale to the system as shown in \cite{Zhao2018,Xu2020} where the dynamic formation and relaxation of a wetting ridge is explored assuming poroelasticity and not just viscoelasticity as done earlier \cite{Shanahan-1st}.

The potential relevance of poroelasticity and of uncrosslinked molecules in a PDMS elastomers in general can be estimated from their volume fraction. While SG184 contains only 4-5\% \cite{Wong2020} of uncrosslinked molecules can this fraction increase up to 60\% for certain PDMS mixtures \cite{PhaseSeparationJPham} and even more for swollen PDMS \cite{PhaseSeparationKJensen,PhaseSeparationJPham,Liu2016}. For colloids \cite{PhaseSeparationKJensen} and droplets \cite{PhaseSeparationJPham} in contact with strongly swollen PDMS, a volume of demixed PDMS oligomers was observed near the TPCL establishing a smooth contact of PDMS and colloid while avoiding the otherwise occurring stress singularities at the TPCL \cite{PhaseSeparationKJensen,PhaseSeparationJPham}.
In \cite{PhaseSeparationJPham} it was even speculated that the phase separation occurs for any swelling ratio but could not be detected due to technical limitations. This is in line with the well known feature that PDMS substrates can restore their hydrophilic properties after oxidation and that a reduced surface tension was observed for water or glycerin droplets that were placed on native PDMS surfaces, which can be interpreted as a thin layer of PDMS oligomers that were extracted from the bulk PDMS and that are cloaking \cite{style2013universal,Zhao2022,PhaseSeparationJPham} or lubricating \cite{Hourlier-Fargette2017,Wong2020} the drops. 
Interestingly, even in articles that neither observed cloaking or mention lubrication explicitly, it was frequently mentioned that water or glycerol drops on PDMS show no or very little hysteresis effects when given sufficient time to relax  \cite{style2013universal,Xu2017,PhaseSeparationKJensen,snoeijer2018paradox,Heyden2021}, which is at least consistent with a lubrication layer, and based on surface energies also cloaking is expected in such a case.  

We therefore conjecture that the demixing observed optically near the TPCL for water droplets on strongly swollen PDMS substrates \cite{PhaseSeparationJPham} can also occur for PS droplets on native PDMS substrates, at the length scale where diffusive and mechanodiffusive effects are balanced. This phenomenon is able to explain the locally increased elastocapillary length, potentially accounting for both the enhanced elevation of the TPCL and the smoother transition of the (cloaked) PS-air interface to the PDMS-air interface observed in experiments compared to theoretical predictions. The hypotheses of poroelasticity and viscoelasticity require further experimental and theoretical investigation to clarify their roles and, if possible, attempt a disambiguation of their respective contributions.

\section*{Acknowledgement}
Funding by the German Research Foundation (DFG) through the projects \#422786086 (LS,RS,KR,BW) and \#422792530 (DP) within the DFG Priority Program SPP 2171 \emph{Dynamic Wetting of Flexible, Adaptive, and Switchable Substrates} is gratefully acknowledged. Furthermore, we acknowledge the help of Hartmut Stadler (Bruker) with NanoIR measurements and of Svenja Pohl  (AG Kickelbick, Inorganic Solid State Chemistry at Saarland University) with ATR-FTIR spectroscopy measurements.  
\bibliographystyle{abbrv}
\bibliography{refs}

\cleardoublepage
\appendix
\section{Appendix / SI}
\setcounter{table}{0}
\setcounter{figure}{0}
\renewcommand{\thefigure}{A\arabic{figure}}
\renewcommand{\thetable}{A\arabic{table}}

\subsection{Characterization of Rheological Properties}


The characteristic parameter describing the viscoelasticity of a material is the (complex) shear modulus $G= G' + iG"$, where
$G'$ and $G"$ are the {storage} and {loss modulus}, respectively. The (visco-) elasticity of the different PDMS substrates was characterized by a Haake-Mars-40 rheometer in plate-plate geometry ($25\,\si{\milli\meter}$ radius). The liquid PDMS mixtures were confined in the shear geometry and cured in place, at 80\,$^{\circ}$C for 90\,minutes, following the same protocol as for the preparation of the PDMS substrates that were used for the dewetting experiments. 
After the preparation of a sample, a frequency sweep test was carried out at room temperature in the range ($\omega = 0.1 - 100$\,rad/s).
For both tested PDMS mixtures, i.e. SG184 and SG186, we found a proportionality between the stress that is applied to the material and its strain response, and therefore a loss modulus $G"$ that is negligible compared the storage modulus $G'$. Consequently one can consider $G= G'$ for SG184 and SG186, which are used here; the obtained results are summarized in the \Cref{tab:params_elast}.
\begin{table}[H]
\centering
\renewcommand{\arraystretch}{1.5}
  \begin{tabular}{lr}
    \hline\hline
    PDMS substrate & $G$ (kPa) \\
    \hline
Sylgard 184 & $595 \pm \,30$ \\
Sylgard 186 & $224 \pm \,30$ \\
    \hline\hline
  \end{tabular}
\caption{Shear modulus $G$ of PDMS rubbers obtained by a Haake-Mars-40 shear rheometer in plate-plate geometry ($25\si{\milli\meter}$ radius, $\omega = 0.1 - 100$\,rad/s) at room temperature. The loss modulus was observed to be negligible, so $G=G'$ for SG184 and SG186.} 
\label{tab:params_elast}
\end{table}


\newpage
\subsection{Droplet Shape Characterisation} \label{AFM-SI}
The topography of the droplets was obtained by atomic force microscopy (AFM) in soft tapping mode, for both top and bottom sides using the lift-off technique explained in the methods section and here in \Cref{fig:lift-off}. 
\definecolor{dropcol}{HTML}{0072A8}
\definecolor{subscol}{HTML}{CDC7B9}
\definecolor{gluecol}{HTML}{49F17D}
\begin{figure*}[ht!]
\centering
a)
\tikzset{grid/.style={gray,very thin,opacity=1}}
\begin{tikzpicture}[scale=0.7,every node/.style={scale=0.7}]   
        \fill[style={thin,subscol}] (0,0,2) -- (0,1,2) -- (4,1,2) -- (4,0,2) -- cycle;
        \fill[style={thin,subscol}] (4,0,0) -- (4,0,2) -- (4,1,2) -- (4,1,0) -- cycle;
        \fill[style={thin,subscol}] (0,1,0) -- (0,1,2) -- (4,1,2) -- (4,1,2) -- (4,1,0) -- (0,1,0);
        \draw[style={thin,subscol!40!black}]  (0,1,0) -- (4,1,0)-- (4,0,0);
	\draw[style={thin,subscol!40!black}] (0,0,2) -- (0,1,2) -- (4,1,2) -- (4,0,2) -- cycle;
        \draw[style={thin,subscol!40!black}] (4,0,0) -- (4,0,2) -- (4,1,2) -- (4,1,0) -- cycle;
        \draw[style={thin,subscol!40!black}] (0,1,0) -- (0,1,2);

        \shade[ball color = dropcol, opacity = 1] (0.5,0.7,-0.2) arc (180:0:0.3);
        \shade[ball color = dropcol, opacity = 1] (0.5,0.7,-0.2) arc (-180:0:0.3 and 0.07);
        \shade[ball color = dropcol, opacity = 1] (1.3,0.4,-0.1) arc (180:0:0.5);
        \shade[ball color = dropcol, opacity = 1] (1.3,0.4,-0.1) arc (-180:0:0.5 and 0.1);
        \shade[ball color = dropcol, opacity = 1] (2.5,0.7,-0.5) arc (180:0:0.2);
        \shade[ball color = dropcol, opacity = 1] (2.5,0.7,-0.5) arc (-180:0:0.2 and 0.04);
        
       \node[] at (1.5,1.8,0) {\large  Equilibrium droplet ensemble};
\end{tikzpicture}
\hfill
b)
\begin{tikzpicture}[scale=0.7,every node/.style={scale=0.7}]  
    \def\xs{0}
    \def\ys{-0.1}
    \fill[color=dropcol]  (-1+\xs,1.2+\ys) arc (180:0:1 and 0.8) (1+\xs,1.2+\ys) arc (0:-180:1 and 0.3) ;
    \fill[color=subscol] 
    (-2.5+\xs,1.07+\ys) arc (-90:-20:1.6 and 0.2) -- 
    (-1+\xs,1.2+\ys) arc (-180:-90:1 and 0.3) -- 
    (\xs,0.9+\ys) arc (-90:0:1 and 0.3) -- 
    (\xs+1,1.2+\ys) arc (-160:-90:1.6 and 0.2) -- 
    (\xs+2.5,1.05+\ys) -- (\xs+2.5,0) -- (-2.5+\xs,0) -- (-2.5+\xs,1.05+\ys) -- cycle; 

    \node[] at (0,0.45) {\large substrate};
    \node[color=white] at (0,1.3) {\large liquid};
    \node[] at (1.6,1.4) {\large air};

    \node[] at (0,2.4) {\large Single droplet cross-section};
    
    \draw[thick,-, black] (\xs+-0.03-3,0) -- (\xs+3,0);
    \foreach \x in {-2.5,-2.4,...,2.5}
   {
       \draw[] (\xs+\x,0) -- (\xs+\x+0.15,-0.15);
   }
\end{tikzpicture}
\hfill
c)
\begin{tikzpicture}[scale=0.7,every node/.style={scale=0.7}]  
    \def\xs{0}
    \def\ys{-0.1}
    \fill[color=dropcol]  (-1+\xs,1.2+\ys) arc (180:0:1 and 0.8) (1+\xs,1.2+\ys) arc (0:-180:1 and 0.3) ;
    \fill[color=subscol] 
    (-2.5+\xs,1.07+\ys) arc (-90:-20:1.6 and 0.2) -- 
    (-1+\xs,1.2+\ys) arc (-180:-90:1 and 0.3) -- 
    (\xs,0.9+\ys) arc (-90:0:1 and 0.3) -- 
    (\xs+1,1.2+\ys) arc (-160:-90:1.6 and 0.2) -- 
    (\xs+2.5,1.05+\ys) -- (\xs+2.5,0) -- (-2.5+\xs,0) -- (-2.5+\xs,1.05+\ys) -- cycle; 

    \draw[ultra thick,-, blue!80!black] (-1+\xs,1.2+\ys) arc (180:0:1 and 0.8); 
    \draw[ultra thick,-, blue!80!black] (-2.5+\xs,1.07+\ys) arc (-90:-20:1.6 and 0.2); 
    \draw[ultra thick,-, blue!80!black] (\xs+1,1.2+\ys) arc (-160:-90:1.6 and 0.2); 

    \node[] at (0,2.4) {\large AFM top scan};
    
    \draw[thick,-, black] (\xs+-0.03-3,0) -- (\xs+3,0);
    \foreach \x in {-2.5,-2.4,...,2.5}
   {
       \draw[] (\xs+\x,0) -- (\xs+\x+0.15,-0.15);
   }
\label{fig:Lift-Off}
\end{tikzpicture}

\vspace{0.2cm}

d)
\tikzset{grid/.style={gray,very thin,opacity=1}}
\begin{tikzpicture}[scale=0.7,every node/.style={scale=0.7}]  
    \def\xs{0}
    \def\ys{-0.1}
    \fill[color=gluecol]  (-2.5+\xs,2.5+\ys) --  (\xs+2.5,2.5+\ys) -- (\xs+2.5,0) -- (-2.5+\xs,0) -- cycle; 
    \fill[color=dropcol]  (-1+\xs,1.2+\ys) arc (180:0:1 and 0.8) (1+\xs,1.2+\ys) arc (0:-180:1 and 0.3) ;
    \fill[color=subscol]  (-2.5+\xs,1.07+\ys) arc (-90:-20:1.6 and 0.2) -- (-1+\xs,1.2+\ys) arc (-180:-90:1 and 0.3) -- (\xs,0.9+\ys) arc (-90:0:1 and 0.3) -- (\xs+1,1.2+\ys) arc (-160:-90:1.6 and 0.2) -- (\xs+2.5,1.05+\ys) -- (\xs+2.5,0) -- (-2.5+\xs,0) -- (-2.5+\xs,1.05+\ys) -- cycle; 

    \node[] at (0,2.1) {\large UV-curable glue};
    
    \draw[thick,-, black] (\xs+-0.03-3,0) -- (\xs+3,0);
    \foreach \x in {-2.5,-2.4,...,2.5}
   {
       \draw[] (\xs+\x,0) -- (\xs+\x+0.15,-0.15);
   }
    \foreach \x in {-2.3,-2.0,...,2.3}
   {
       \draw[thick,yellow!50!red,->] (\xs+\x,2.9) -- (\xs+\x,2.5);
   }
   \node[yellow!50!red] at (0,3.2) {\large UV light};
\end{tikzpicture}
\hfill
e)
\begin{tikzpicture}[scale=0.7,every node/.style={scale=0.7}]  
    \def\xs{0}
    \def\ys{-0.2}
    
    \fill[color=gluecol,rotate=180]  (-2.5+\xs,2.35+\ys+0.1) --  (\xs+2.5,2.35+\ys+0.1) -- (\xs+2.5,0.9) -- (-2.5+\xs,0.9) -- cycle; 
    \fill[color=dropcol,rotate=180]  (-1+\xs,1.2+\ys) arc (180:0:1 and 0.8) (1+\xs,1.2+\ys) arc (0:-180:1 and 0.3) ;
    \fill[color=white,rotate=180]  (-2.5+\xs,1.08+\ys) arc (-90:-20:1.6 and 0.2) -- (-1+\xs,1.2+\ys) arc (-180:-90:1 and 0.3) -- (\xs,0.9+\ys) arc (-90:0:1 and 0.3) -- (\xs+1,1.2+\ys) arc (-160:-90:1.6 and 0.2) -- (\xs+2.5,1.05+\ys) -- (\xs+2.5,0) -- (-2.5+\xs,0) -- (-2.5+\xs,1.08+\ys) -- cycle; 

    \draw[ultra thick,-, red,rotate=180]  (-2.5+\xs,1.08+\ys) arc (-90:-20:1.6 and 0.2) -- (-1+\xs,1.2+\ys) arc (-180:-90:1 and 0.3) -- (\xs,0.9+\ys) arc (-90:0:1 and 0.3) -- (\xs+1,1.2+\ys) arc (-160:-90:1.6 and 0.2); 
    
    \draw[thick,-, black] (\xs+-0.03-3,-2.25) -- (\xs+3,-2.25);
    
    \node[] at (0,-2.02) {\large UV-curable glue};
    \draw[->,double,black] (-0.45,-0.5) arc (180:0:0.45) ;
    \node[] at (0,0.3) {\large AFM bottom scan};
\end{tikzpicture}
\hfill
f)
\begin{tikzpicture}[scale=0.7,every node/.style={scale=0.7}]  
\def\xs{0}
\def\ys{-0.1}
\def\Hh{3.5}
\def\HH{3.4}
    \fill[line width=2.6, color=black!30!white]  (-1.5,3) --  (1.5,3) -- (1.5,1.2) -- (-1.5,1.2) -- cycle; 
    \draw[ultra thick, color=black!30!white]  (-1.5,3) --  (1.5,3) -- (1.5,1.2) -- (-1.5,1.2) -- cycle; 
    \fill[ultra thick, color=black!10!white]  (-1.5,3.5) --  (1.5,3.5) -- (1.5,1.5) -- (-1.5,1.5) -- cycle;
    \draw[line width=2.6, color=black]  (-1.5,3.5) --  (1.5,3.5) -- (1.5,1.5) -- (-1.5,1.5) -- cycle;
    
    \draw[line width=3, color=black!30!white]  (-0.2,1.1) -- (0.2,1.1); 
    \draw[line width=2.3, color=black!30!white]  (-0.5,1) -- (0.5,1); 
    \draw[line width=2.3, color=black!30!white]  (-0.5,1) -- (0.5,1); 
    \draw[thick,-, white] (-0.03-3,0.7) -- (3,0.7);
    
    \node at (0,1.3) [circle,fill,inner sep=1.5pt]{};
    \node[] at (0,3.2) {\small\textsf{3D AFM SCAN}};
    \draw[thick,-, blue!80!black, scale=0.5] (-1+\xs,1.2+\ys+\Hh) arc (180:0:1 and 0.8); 
    \draw[thick,-, blue!80!black, scale=0.5] (-2.5+\xs,1.07+\ys+\Hh) arc (-90:-20:1.6 and 0.2); 
    \draw[thick,-, blue!80!black, scale=0.5] (\xs+1,1.2+\ys+\Hh) arc (-160:-90:1.6 and 0.2); 
    \draw[thick,-, red, scale=0.5]  (-2.5+\xs,1.08+\ys+\HH) arc (-90:-20:1.6 and 0.2) -- (-1+\xs,1.2+\ys+\HH) arc (-180:-90:1 and 0.3) -- (\xs,0.9+\ys+\HH) arc (-90:0:1 and 0.3) -- (\xs+1,1.2+\ys+\HH) arc (-160:-90:1.6 and 0.2);
\end{tikzpicture}
\caption{Sketch of experimental lift off process: a) Preparation of equilibrium droplet ensemble, b) consideration of cross-sections of single droplets, c) AFM top scan of PS/air and PDMS/air interface, d) covering the sample by UV-curable glue, e) peeling off the glue/PS from the PDMS substrate, flipping of the sample, and measurement of AFM bottom scan of formerly PS/PDMS and PDMS/air interface, f) composition of top and bottom AFM scan to a complete 3D droplet with postprocessing software.}
\label{fig:lift-off}
\end{figure*}
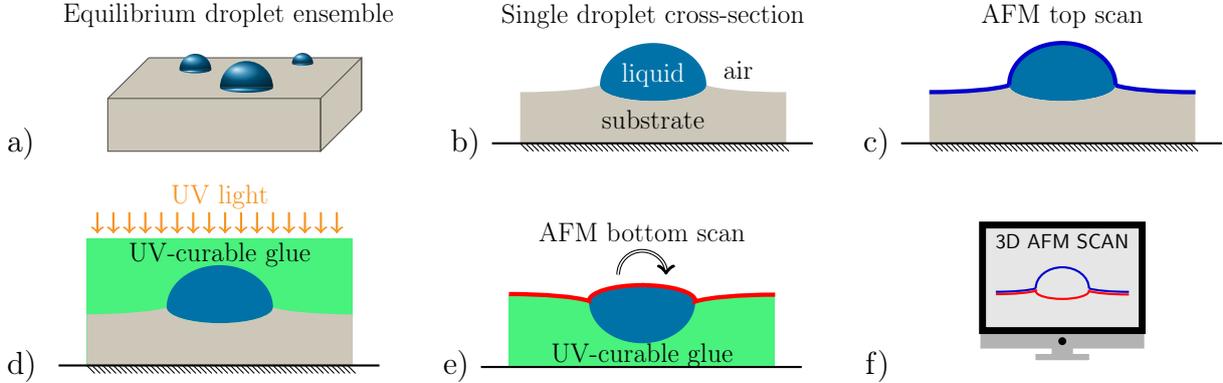

\begin{figure}[b!]
   \centering
   \includegraphics[width=0.6\textwidth]{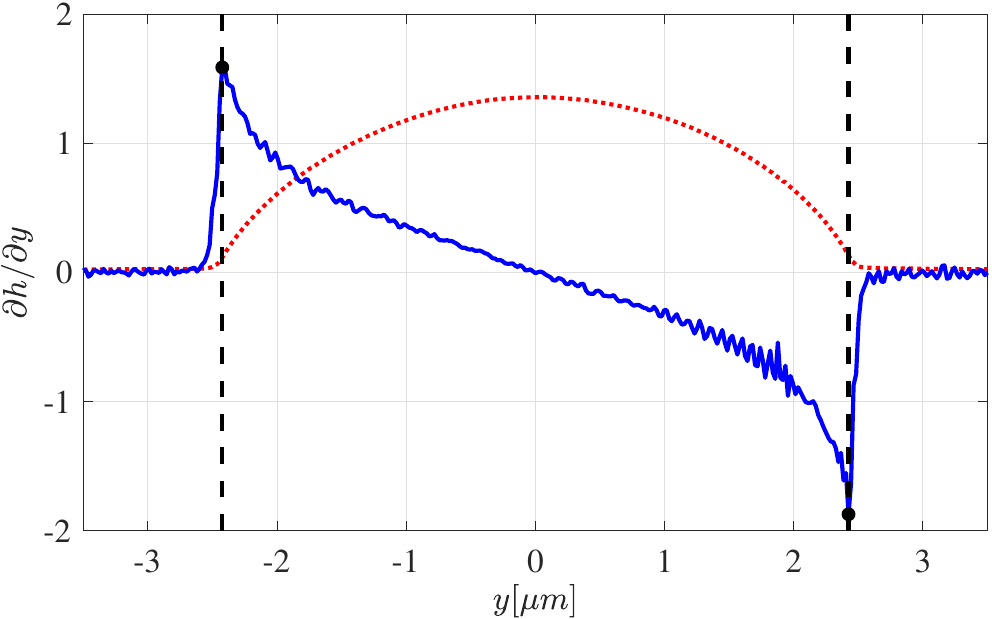}
   \caption{\label{fig: Droplet-SG184_1stderivative} Droplet cross-section on SG184 obtained with AFM (red dots) with corresponding 1st derivative (blue solid). The three-phase contact line can be clearly identified by the sharp local minimum, respectively maximum in the 1st derivative.}
   \label{fig:si_contour_derivative}
\end{figure}

\begin{figure}[ht!]
{\includegraphics[height=0.24\textwidth]{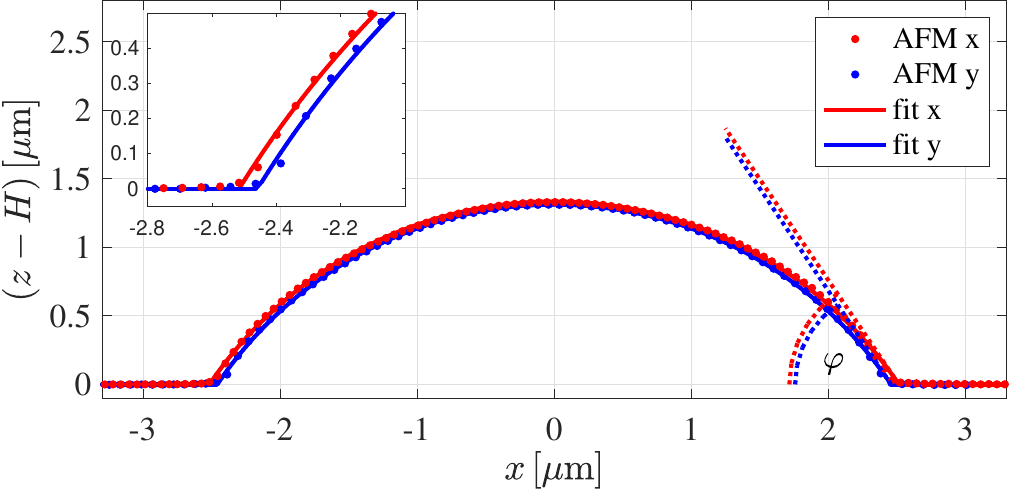}}
\hfill%
{\includegraphics[height=0.24\textwidth]{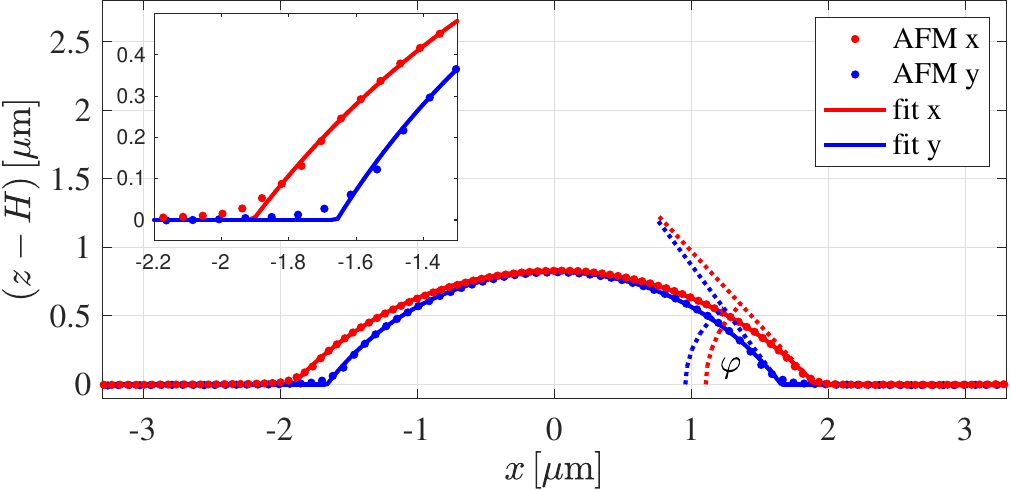}}
\caption{Circular arc fits to AFM cross sections in two perpendicular directions for droplets on (left) SG184 and (right) SG186 PDMS substrates, where Young contact angles $\theta$ are indicated.}
\label{fig:Spherical-cap-fit}
\end{figure}

The three-phase contact line (TPCL) was identified by contour analysis of the AFM profile, see \Cref{fig:si_contour_derivative}. 
The basis of this contour analysis is that the free PS-air surface has a constant curvature, and can be fitted by a circular arc, as shown in \Cref{fig:Spherical-cap-fit}. However, since the second derivative of point data, such as an AFM height contour, is extremely noisy, we instead use the first derivative from the AFM scans, where a sudden change in monotonicity indicates the exact position of the TPCL in the cross-section, as shown in \Cref{fig:si_contour_derivative}. 
One step of smoothing, involving the nearest two scanlines, was applied to slightly reduce the noise in the first derivative, allowing the resulting position of the TPCL to be determined with an accuracy of $\pm 1\,\textrm{px}$ relative to the AFM scan resolution.
The determination of the TPCL using this strategy was proven to be very robust and was also used on cross sections of all the presented samples.
Alternatively, it would be also possible in principle to determine the TPCL via the phase contrast of the AFM, which indicates a material contrast. However, the phase contrast on those samples imaged in soft tapping is not very robust and precise, e.g. due to slight contamination of the AFM tip, and is thus not used here. 

\begin{figure}[b!]
\centering
{\includegraphics[height=0.44\textwidth,trim={2.8cm 0 3.3cm 0.7cm},clip]{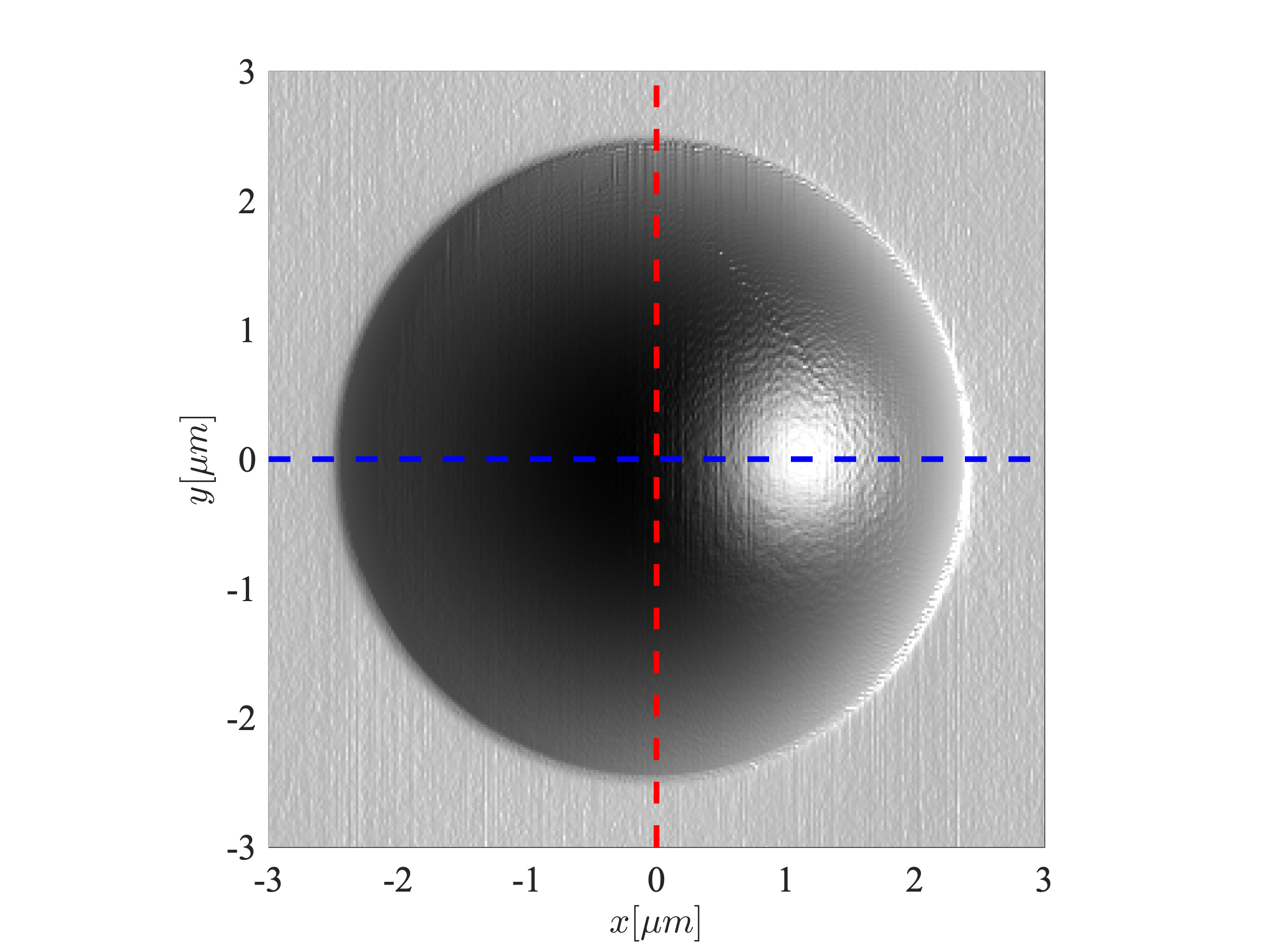}}\qquad%
{\includegraphics[height=0.44\textwidth,trim={3.5cm 0 3.3cm 0.7cm},clip]{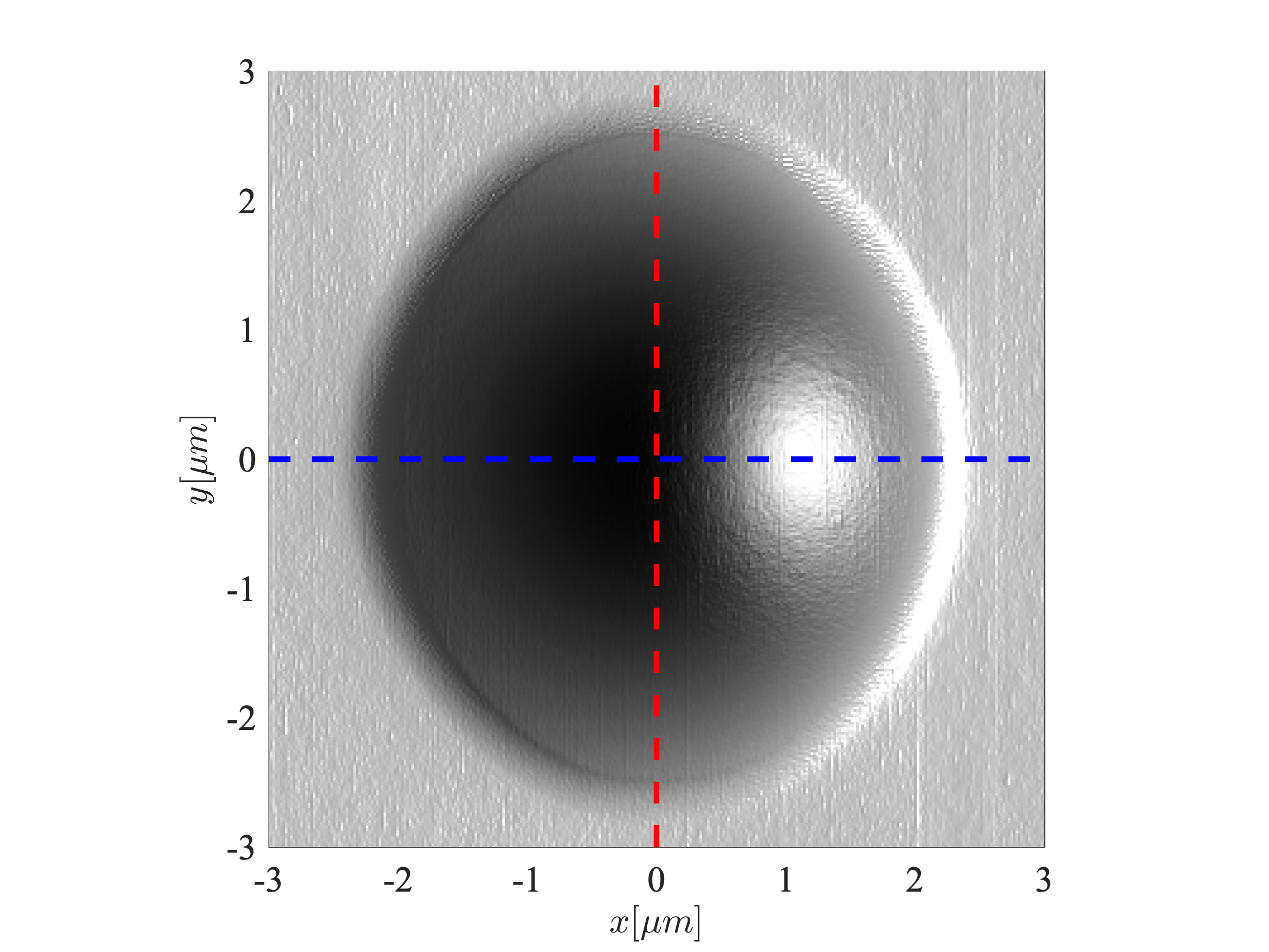}}
\caption{AFM topside scans of PS droplets on (left) SG184 and (right) SG186 substrates that were shown in \Cref{fig:Spherical-cap-fit}. The PS droplet on SG184 exhibits rotational symmetry, whereas the droplet on SG186 shows a noticeable ellipticity. Gray shading based on height and lighting based on slope to improve visibility of droplet shape. The dashed lines indicate the cross-sections used in \Cref{fig:Spherical-cap-fit}.}
\label{fig:AFM_TopView}
\end{figure}

AFM images of polystyrene (PS) droplets on SG184 and 186 substrates are shown in \Cref{fig:Spherical-cap-fit} together with circular arc fits along the smallest and largest axis of the drop. 
The droplet radius on SG184 varies by $3\%$, i.e. within the accuracy of the AFM, whereas the drop radii on SG186 vary by $13\%$ between both measurement directions indicating an elliptical drop shape that is not fully equilibrated but also does not equilibrate further on experimentally achievable time scales, as described in more detail in the main text. Accordingly, the contact angle on the well equilibrated PS droplets of SG184, $\theta=(55.9\pm 0.3)\si{\degree}$ has only a small uncertainty. For PS droplets on SG186, instead, we observe stronger deviations from a axisymmetrical droplet shape, so that the contact angle for the largest droplets vary between $\theta=47.0\si{\degree}$ (long axis) and $\theta=52.7\si{\degree}$ (short axis).


\subsection{Surface Chemistry Characterization} \label{Chemical-Composition}
\begin{figure}[ht!]
\centering
\includegraphics[width=0.49\textwidth,trim={1cm 0cm 2.5cm 0cm},clip]{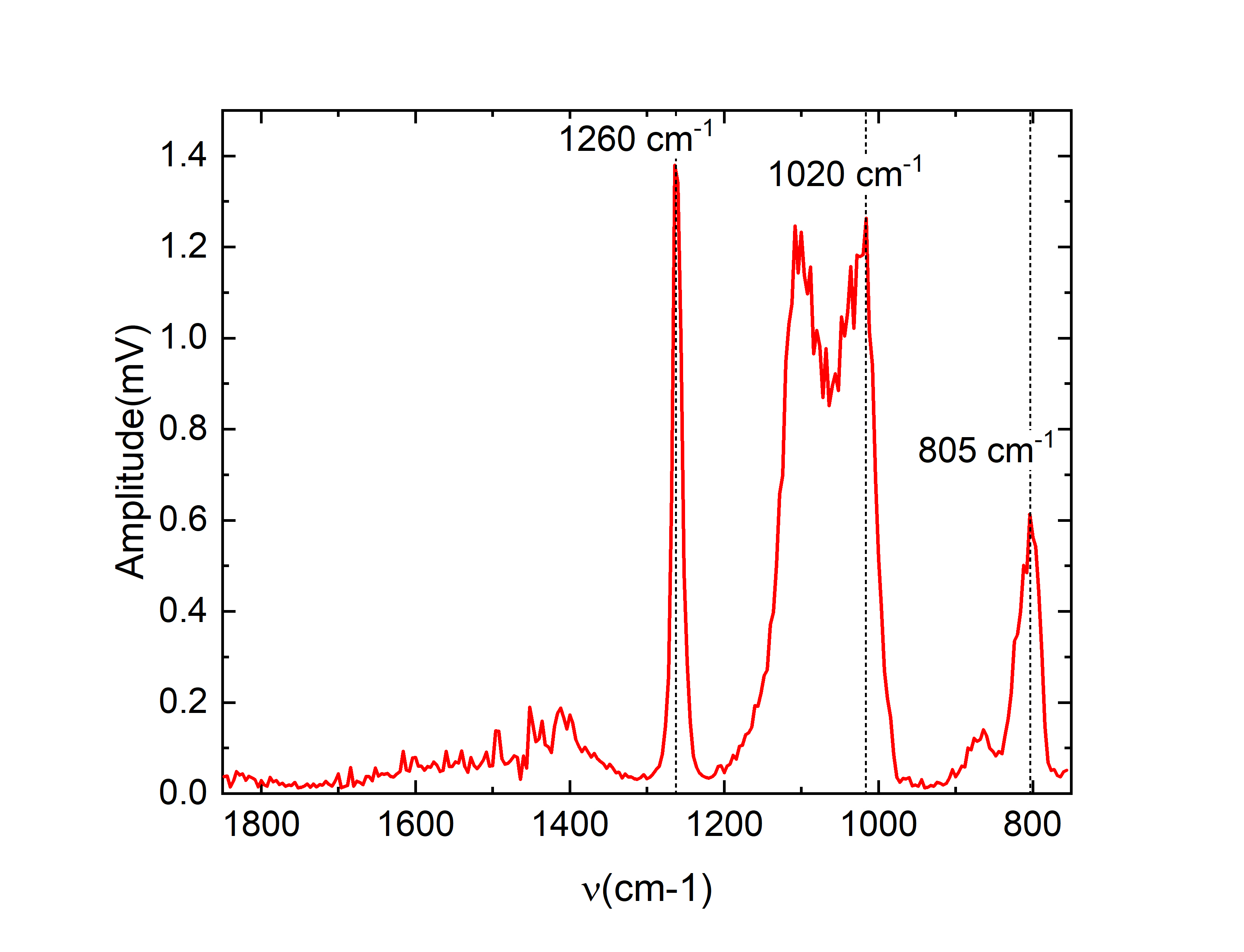}%
\includegraphics[width=0.49\textwidth,trim={1cm 0cm 2.5cm 0cm},clip]{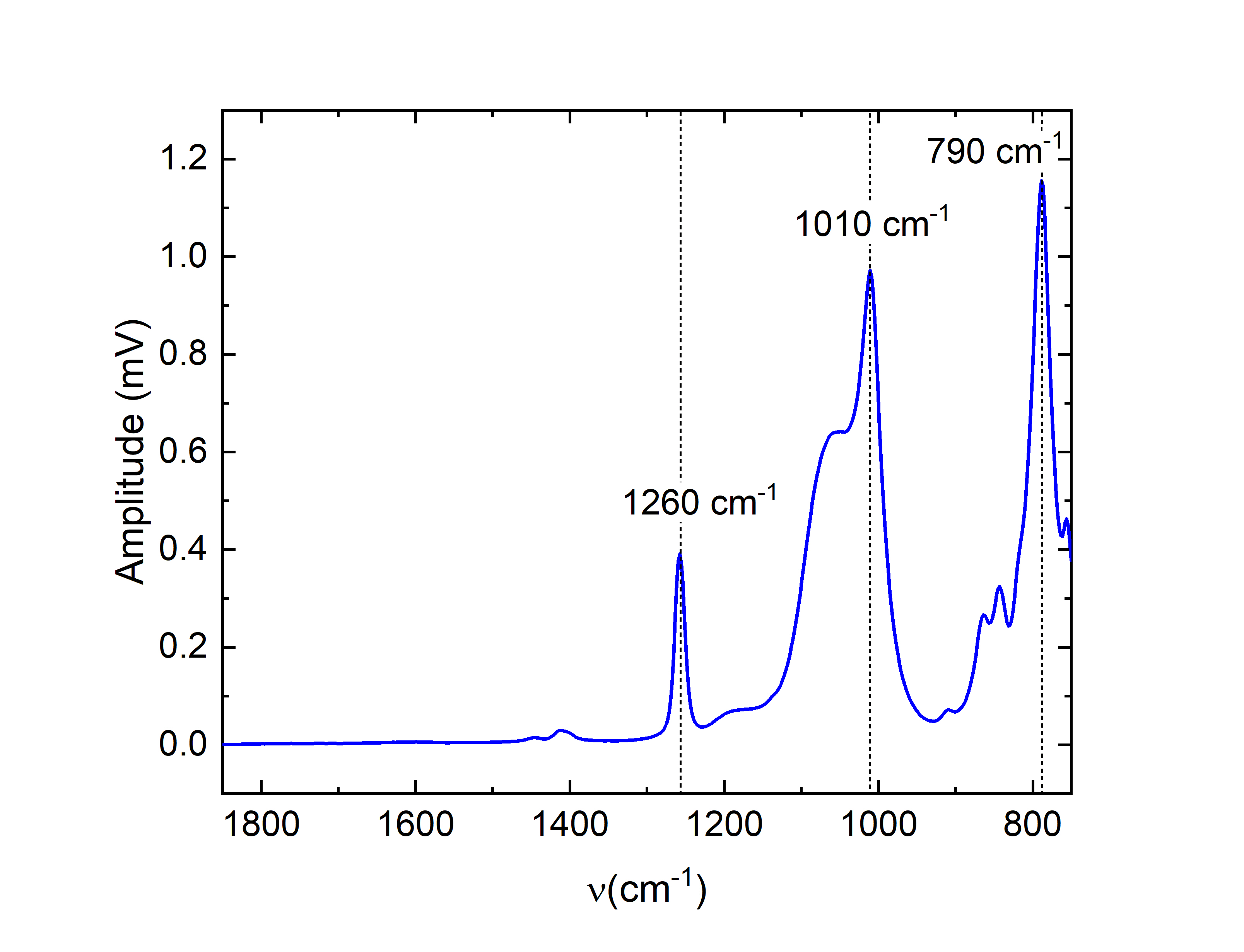}
\caption{\label{fig:ATR-FTIR_Crosslinked_SG184} (left) NanoIR (Bruker) absorption spectrum measured on top of a PS droplet showing the presence of PDMS peaks. (right) IR absorption spectrum Obtained by ATR-FTIR for crosslinked SG184.} 
\end{figure}

Section \ref{AFM-SI} shows how the topography signals obtained by AFM allow to precisely identify the contours of PS droplets. However, this technique is based on the physical response of the material and does not give precise information about the chemical composition of the probed surface. 
To explore the chemical composition at the surface of the PS droplets, \textbf{NanoIR} (Bruker) was applied, \Cref{fig:ATR-FTIR_Crosslinked_SG184}\,(left). The NanoIR technique is based on a pulsed and tunable IR laser focused on the sample and synchronized with the AFM tapping frequency. When the applied wavelength matches an absorbance band of the substrate, it causes a local thermal expansion of the surface that is detected by the AFM tip. By this strategy, spatially resolved infrared absorption spectra are obtained that can be correlated with the presence of specific molecular bands, giving precise information about the molecular composition of the probed surface.
However, the surface sensitivity of NanoIR is not precisely known, but we can assume that the dominant signal comes from a depth below $100\,\si{\nano\meter}$.  

In \Cref{fig:ATR-FTIR_Crosslinked_SG184}\,(right), the infrared spectra of a crosslinked bulk SG184 sample obtained by attenuated infrared spectroscopy (ATR-FTIR) is shown for comparison. Both spectra show three peaks that are characteristic for PDMS (a peak at $1260~\si{\centi\meter}^{-1}$ for the Si--(CH$_3$)$_3$ band, at $805~\si{\centi\meter}^{-1}$ for Si--CH$_3$ and at $1020~\si{\centi\meter}^{-1}$ for Si--O--Si) and thus reveal the presence of a thin layer of PDMS on top of the PS droplet. This observation is in line with expectations from the positive spreading coefficient favoring the cloaking of the PS-droplet by PDMS in order to lower its surface tension. However, slight differences can be observed between both spectra, for example the relative height of the peaks at $805~\si{\centi\meter}^{-1}$ and at $1260~\si{\centi\meter}^{-1}$, are inverted in the two spectra. This difference can probably be explained by the fact that the bulk PDMS has a higher crosslinking density than the liquid PDMS that migrates to top of the PS droplet, which can be correlated to a higher percentage of Si--CH$_3$ groups for the liquid PDMS and a higher Si--(CH$_3$)$_3$ groups for the bulk PDMS. 

\cleardoublepage

\subsection{Axisymmetric Droplet Relaxation Model}
\label{sec:axisymmetric_discrete}
Below, we detail the space and time discretization for the sharp-interface model that describes droplet relaxation via the deformation $\chi(t)$ as $t \to \infty$. This algorithm's implementation achieves the necessary robustness and precision to predict stationary axisymmetric droplets for a range of droplet radii relevant to the capillary length. The initial shape of the domain $\Omega^0$ is illustrated in \Cref{fig:sketch}.

Assuming axisymmetry, we replace three-dimensional volume and surface measures by $\mathrm{d}\bs{x}=2\pi r\,\mathrm{d}r\,\mathrm{d}z$ and 
$\mathrm{d}\bs{a}=2\pi r\,\mathrm{d}s$ and define the axisymmetric deformation gradient  
\begin{align*}
\bs{F}=\nabla\bs{\chi}:=\begin{pmatrix} \partial_r \chi_r & 0 & \partial_z \chi_r \\ 0 & r^{-1} \chi_r  & 0 \\ \partial_r \chi_z & 0 & \partial_z \chi_z \end{pmatrix}
\end{align*}
and rewrite the energy \eqref{equ:sharp_lagrange_energy} and the evolution in \eqref{eqn:extgrad_si}  correspondingly. Following \cite{schmeller2023gradient,schmeller2023sharp},  we discretize \eqref{eqn:extgrad_si} by using P$_2$ finite elements for the deformation $\chi$ and use an incremental minimization strategy to discretize in time and ensure discrete energy descent. The boundaries of the subdomains $\Omega_i$ are assumed polygonal and resolved by the edges of the computational triangular mesh $\Omega_h=\cup_{n=1}^{N_\textrm{element}} T_h$. Near the contact line we employ a heuristic spatial refinement procedure in order to resolve singular elastic deformations due to the capillary ridge at the TPCL caused by the interfacial forces. The implementation of the nonlinear problem is provided in FEniCS \cite{logg2012automated}.

\begin{figure}[hb!]
    \centering
    \includegraphics[width=0.66\textwidth]{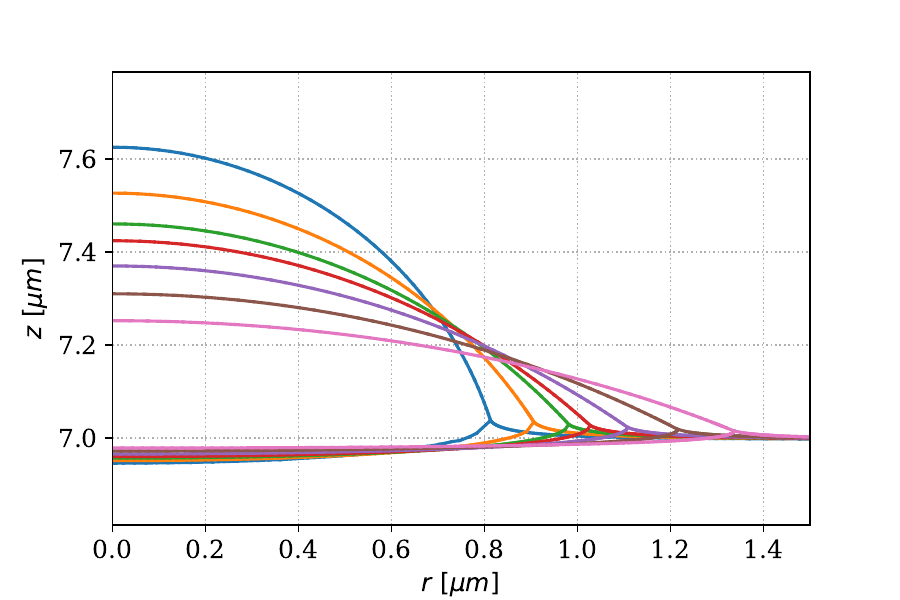}

    \caption{Stationary droplet shapes for different initial radii $R=r_x$ corresponding to \Cref{fig:energy_minima}. Note that the red full lines corresponds to the energy minimizer in \Cref{fig:energy_minima} and corresponds approximately to the parameters of the middle droplet on SG186 in \Cref{fig:comp_exp_theo}.}
    \label{fig:shape_minima}
\end{figure}

\begin{figure}[ht!]
    \centering
    \includegraphics[width=0.999\textwidth]{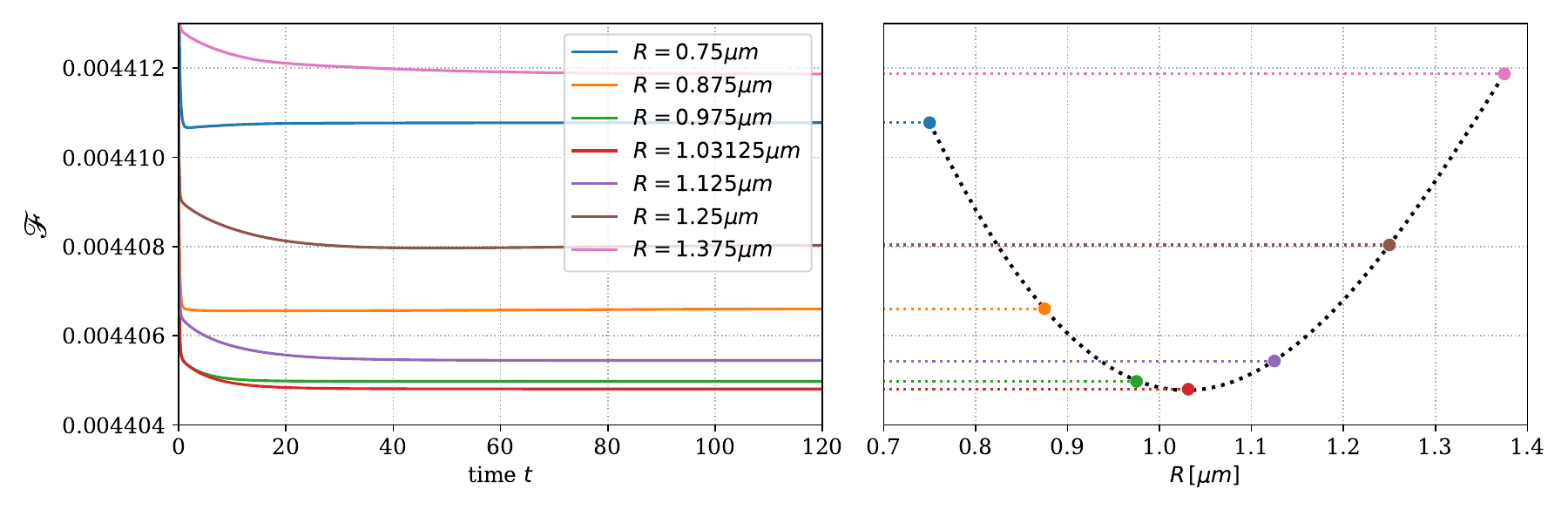}
    \caption{(left) Energy evolution $\mathcal{F}(t)$ of droplet configurations approaching stationary states for different initial radii $R=r_x$ and (right) corresponding stationary energies as a function of droplet radius $R$. The black dotted line is a fitted 4th order polynomial.}
    \label{fig:energy_minima}
\end{figure}

In the dynamic relaxation model, we impose no-slip boundary conditions, i.e. the displacements and velocities along the substrate-liquid interface are continuous. This leads to a pinning of the contact line in the sense that material points on each side close to the PS-PDMS interface move jointly and the displacement of the contact line also generates a (singular) elastic energy. Therefore, the equilibrium state depends on the chosen initial data and does not minimize the free energy in the space of all admissible shapes, e.g., see the different stationary droplet shapes in \Cref{fig:shape_minima}.

To achieve the global minimal energy for an axisymmetric droplet, we vary the initial shape of the liquid domain $\Omega^0_\ell$ and therefore the position of the initial contact line by choosing the initial radius $r_x$ so that $r_z=r_z(V)$ is determined by the given droplet volume $V$. To match an individual AFM measurement, the droplet volume $V$ is determined from the experimental data and we obtain the values stated in \Cref{tab:droplet_data}. We vary the initial radius $r_x$, compute the resulting stationary shape as $t\to\infty$ in the left panel of \Cref{fig:energy_minima}, and compute its equilibrium free energy as a function of $r_x$ as shown in the right panel of \Cref{fig:energy_minima}. Usually we compute the energy of stationary states only for a few $r_x$ values, e.g., 7 values in \Cref{fig:energy_minima}, and find the minimum by interpolating with a polynomial. This optimal radius $r_x$ is used to compute the optimized droplet shape -- this is the shape to which the dynamics would have converged with the more admissible boundary condition at the PS-PDMS interface.

\begin{table*}[ht!]
\centering
\renewcommand{\arraystretch}{1.5}
  \begin{tabular}{lcc}
    \hline\hline
    PDMS & volume $[\si{\cubic\micro\meter}]$ & radius $[\si{\micro\meter}]$\\
    \hline
    \hline
SG184 & $20.8 \pm 0.5$ & 2.804\\
SG184 & $2.16 \pm 0.02$ & 1.314 \\
SG184 & $0.147 \pm 0.008$ & 0.512\\
SG186 & $15.1 \pm 0.1$ $(10.61\pm 0.1)$ & 2.676\\
SG186 & $0.85 \pm 0.01$ $(0.60\pm 0.04)$ & 0.975\\
SG186 & $0.042 \pm 0.002$ $(0.027\pm 0.005)$ & 0.338 \\
    \hline\hline
  \end{tabular}
\caption{Computed volumes and radius of (assumed axisymmetric) PS droplets shown in \Cref{fig:comp_exp_theo}. Values in brackets for SG186 are based on short axis.}
\label{tab:droplet_data}
\end{table*}

\newpage
\subsection{Dependence on Solid Angle}
\label{sec:theta_S}
\begin{figure}[H]
    \centering
    \includegraphics[width=0.65\textwidth]{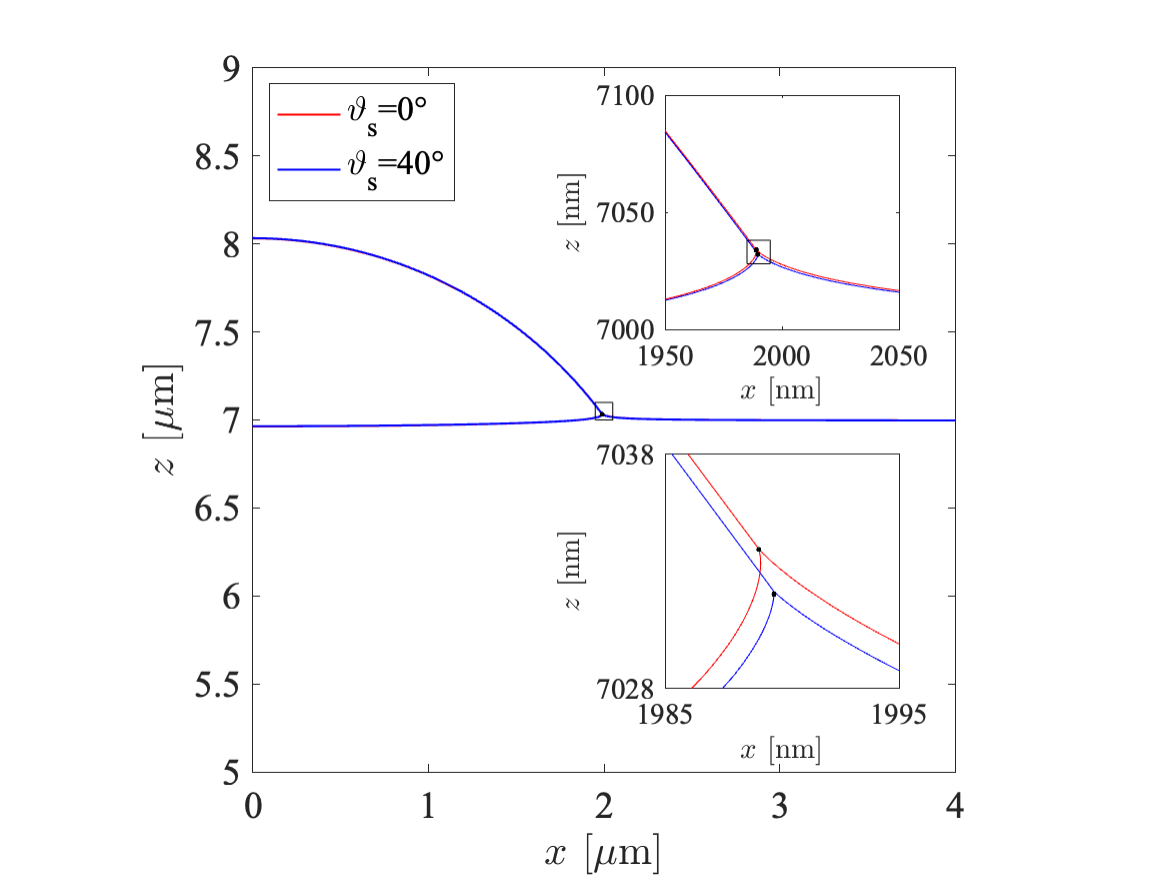}
    \caption{Comparison of numerical solutions for solutions of sharp-interface model with surface tensions with the same Young angle but different solid opening angles $\vartheta_s$.}
    \label{fig:sub_angles}
\end{figure}

In this work, we employ the hybrid construction \eqref{eq:hybrid} to obtain surface tensions compatible with the observation of cloaking and the Young angle for large droplet sizes. However, it has been noted in the literature that, particularly for small or vanishing solid angles, defining and measuring the Neumann angles \(\vartheta_\ell, \vartheta_s\) can be somewhat problematic due to the singular nature of the elastocapillary ridge \cite{pandey2020singular}. We verify and confirm this observation by showing in \Cref{fig:sub_angles} droplets with radii \(R \approx 2\,\si{\micro\meter}\) and surface tensions corresponding to the same Young angle but two solid angles, \(\vartheta_\text{s} = 0\si{\degree}\) and \(\vartheta_\text{s} = 40\si{\degree}\). On the droplet scale, the interface shapes appear identical, and only at scales of \(\pm\,5\,\si{\nano\meter}\) are small deviations in angle and height observable, well below the resolution of realistic AFM measurements. 

This leads us to conclude that, in the moderately soft limit for \(R \ll \lambdacap\) and small opening angle \(\vartheta_\text{s}\), the Neumann angle can only be observed at scales smaller than the elastocapillary length $\lambdacap$ and potentially even at or below molecular scales $a\sim 10^{-9}\,\si{\meter}$.

\subsection{Dependence on Substrate Thickness}
\label{sec:dependence_on_H}
\begin{figure}[H]
    \centering
    \includegraphics[width=0.75\textwidth]{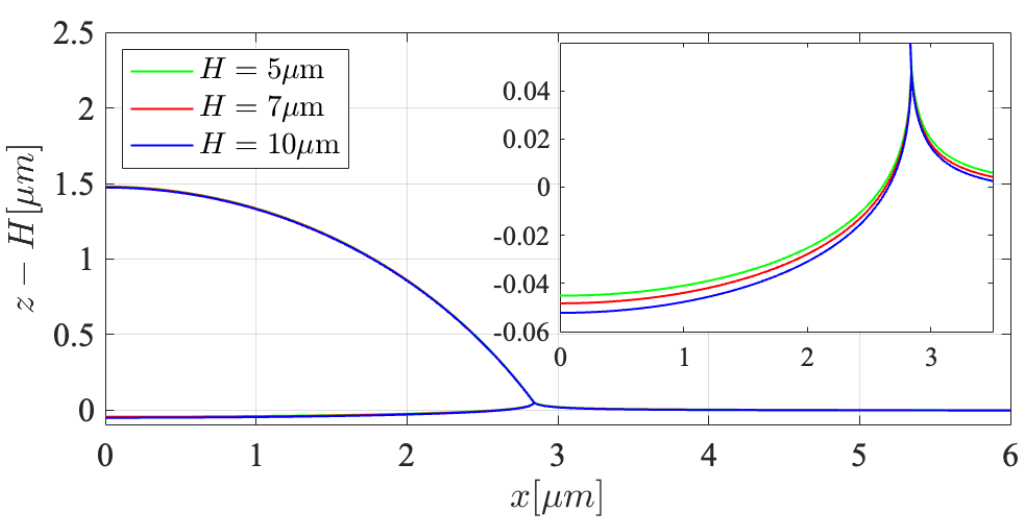}
    \caption{Numerical comparison of three substrate heights, where the base height is subtracted from the $z$ coordinate.}
    \label{fig:sub_heigth}
\end{figure}

The experiments and simulations presented in this paper were performed for moderately soft droplets \(R \ll \lambdacap\) on moderately thick substrates, where the radius of the largest droplets is comparable to the substrate thickness \(H = 7\,\si{\micro\meter}\). To estimate the effect of substrate thickness, we conducted a small study on such a droplet, see \Cref{fig:sub_heigth}. The inset of the figure shows that, within the range \(H = 5\text{--}10\,\si{\micro\meter}\), the indentation depth at the PS-PDMS interface, approximately \(50\,\si{\nano\meter}\), changes only by about \(\pm 5\,\si{\nano\meter}\). This variation is comparable to the experimental interface roughness. Furthermore, the position and elevation of the TPCL and the wetting ridge are not visibly affected by the substrate thickness. Therefore, variations in substrate thickness do not account for the observed discrepancies near the TPCL.

\subsection{Magnitude of displacement near TPCL} 
\label{sec:local_shear_TPCL}
\begin{figure}[h]
\centering
\includegraphics[width=.75\textwidth]{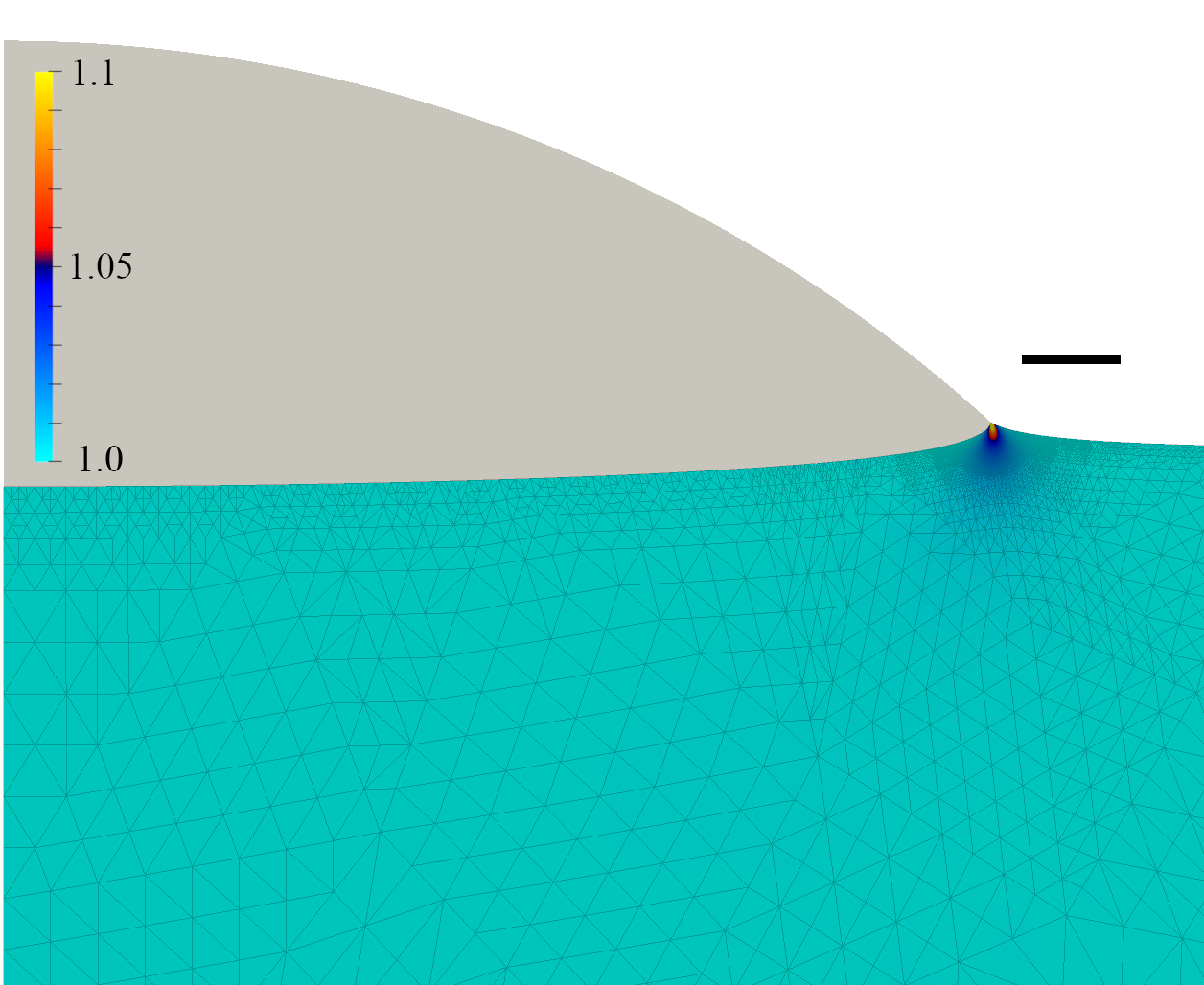}
\caption{For a micrometer sized droplet we show $\sqrt{\mathrm{tr}{(\boldsymbol{F}^T\boldsymbol{F})}/3}$ to indicate the relative local stretching near the TPCL. The black bar indicates 100nm length scale.}
\label{Fig:nonlin_elast}
\end{figure}

In order to estimate the potential impact of the Shuttleworth effect, we display \(\sqrt{\text{tr}(\bs{F}^T\bs{F})/3}\) as a simple local measure of stretching to estimate the size of spatial regions where surface tension and energy could potentially deviate from each other. The shown numerical solution in \Cref{Fig:nonlin_elast} corresponds to a droplet with radius \(R \sim 1\,\si{\micro\meter}\), where the back bar indicates \(100\,\si{\nano\meter}\), which is comparable to the elastocapillary length \(\lambdacap = 80\,\si{\nano\meter}\) for SG186. The region where the local stretching exceeds \(10\%\), highlighted by lighter colors near the TPCL, is clearly much smaller than \(\lambdacap\) and restricted to a scale of about \(10\,\si{\nano\meter}\).

\subsection{Dependence on PDMS Shear Modulus}
\label{sec:dependence_G}
\begin{figure*}[ht!]
\centering
\includegraphics[width=0.5\textwidth]{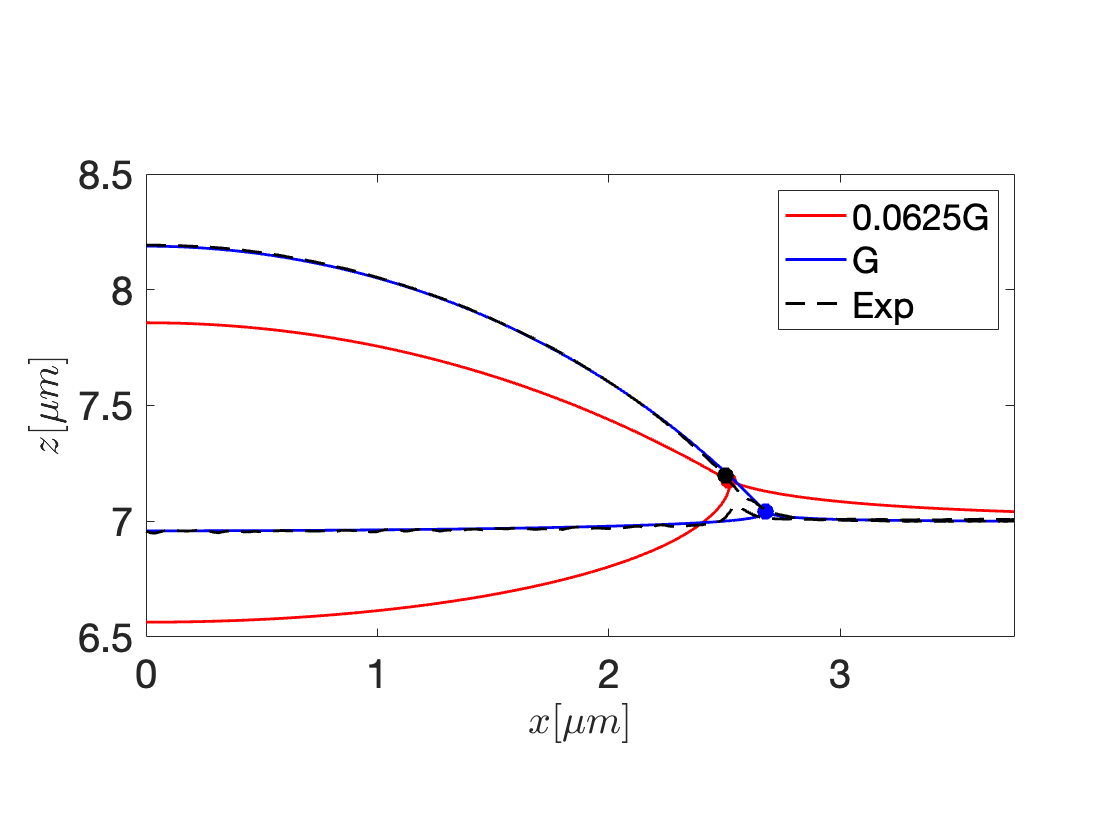}\hfill
\includegraphics[width=0.5\textwidth]{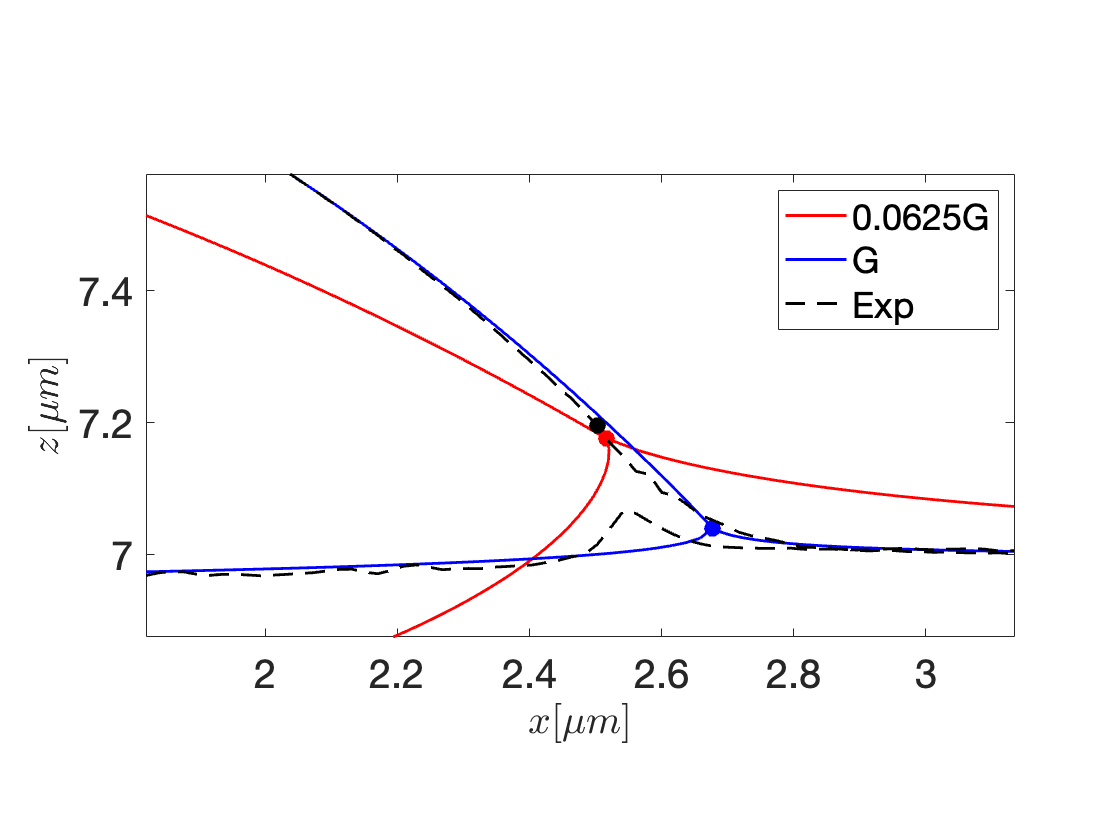}
\caption{Comparison of experimental AFM cross section for SG186 (black dashed line) and position of contact line (black dot) compared to theoretical predictions,  global shape (left), about three times magnified (right). The numerical shape and TPCL are shown in blue for the correct shear modulus $G=\Gsoft$, while the data in red color display the shape computed for much softer PDMS substrate with shear modulus $\tfrac{1}{16} \Gsoft^\text{exp}$.}
\label{fig:comp_G_Sy186}
\end{figure*}

In the manuscript, it is argued that the observed discrepancies, particularly the enhanced elevation of the TPCL, can only be explained by a locally enhanced elastocapillary length. Since we also argue that a global variation in the shear modulus is necessary to achieve good agreement with global droplet shapes, one might consider the possibility of matching shapes using a globally increased elastocapillary length $\lambdacap$. In \Cref{fig:comp_G_Sy186}, we compare theoretical and experimental profiles for a drastically reduced PDMS shear modulus in SG186 $\tfrac{1}{16} \Gsoft^\text{exp}$. While this reduction predicts a similar elevation of the TPCL as observed in the experiment, the theoretical global droplet profile deviates significantly from the experimentally measured profile, thereby excluding the possibility of global variations in the shear modulus as an explanation.

\cleardoublepage

\end{document}